\documentclass[11pt]{article}
\usepackage{epsfig}
\usepackage{amssymb}
\usepackage[]{times}

\makeatletter \@addtoreset{equation}{section}

\makeatother
\newcommand{\be}{\begin{equation}}
\newcommand{\ee}{\end{equation}}
\newcommand{\bea}{\begin{eqnarray}}
\newcommand{\eea}{\end{eqnarray}}

\begin{document}
\title{Stability of Axisymmetric Pendular Rings}
\author{Leonid G. Fel$^1$
 and Boris Y. Rubinstein$^2$\\ \\
$^1$Department of Civil and Environmental Engineering,\\
Technion -- Israel Institute of Technology, Haifa, 32000, Israel\\
$^2$Stowers Institute for Medical Research, 1000 E 50th St,
Kansas City, MO 64110, USA}
\date{}
\maketitle
\begin{abstract}
Based on the Weierstrass representation of second variation we develop a
non-spectral theory of stability for isoperimetric problem with minimized and
constrained two-dimensional functionals of general type and free endpoints
allowed to move along two given planar curves. We apply this theory to the
axisymmetric pendular ring between two solid bodies without gravity to determine
the stability of menisci with free contact lines. For catenoid and cylinder
menisci and different solid shapes we determine the stability domain. The other
menisci (unduloid, nodoid and sphere) are considered in a simple setup between
two plates. We find the existence conditions of stable unduloid menisci with
and without inflection points.
\end{abstract}
\section{Introduction}\label{s1}
A capillary surface is an interface separating two non-mixing fluids adjacent to
each other. Its shape depends on liquid volume and on boundary conditions (BC)
specified at the contact line (CL) where the liquids touch the solid. A pendular
ring (PR) is one of the well studied among different types of drops (sessile,
pendant \cite{Wente1980}, {\it etc}.). It emerges when a small amount of fluid
forms a axisymmetric liquid bridge with interface (meniscus) between two
axisymmetric solids. A history of the PR problem without gravity shows a
remarkable interaction between theoretical physics and pure mathematics and can
be traced in two directions: evolution of menisci shapes (including their
volume $V$, surface area $S$ and surface curvature $H$ calculation) and study
of their stability.

Delaunay \cite{Delaunay1841} was the first who classified all surfaces of
revolution with constant mean curvature (CMC) in his study of the Young-Laplace 
equation (YLE). These are cylinder (${\sf Cyl}$), sphere (${\sf Sph}$), catenoid
(${\sf Cat}$), nodoid (${\sf Nod}$) and unduloid (${\sf Und}$). Later Beer 
\cite{Beer1857} gave analytical solutions of YLE in elliptic integrals and 
Plateau \cite{Pl1873} supported the theory by experimental observations. For a 
whole century almost no rigorous results were reported on the computation of 
$H$, $V$ and $S$ of PR. In the 1970s Orr {\it et. al.} \cite{Orr1975} gave such 
formulas for all meniscus types in case of a solid sphere contacting a solid 
plate. A new insight into the problem was presented recently in \cite{RF13} for
the case of separated solid sphere and plate as a nonlinear eigenvalue equation
with a discrete spectrum. The existence of multiple solutions of YLE for given 
PR volume reported in \cite{RF13} poses a question of local stability of 
menisci.

The first step toward the modern theory of PR stability was made by Sturm
\cite{Sturm1841} in appendix to \cite{Delaunay1841}, characterizing CMC surfaces
as the solutions to isoperimetric problem (${\bf IP}$). Such relationship
between a second order differential equation and a functional reaching its
extremal value was known at the time. The basis of calculus of variations was
laid in the 1870s by Weierstrass in his unpublished lectures \cite{Weier1927}
and extended by Bolza \cite{Bolz1904} and others. The difficult part of the
theory deals with the second variation in vicinity of extremal solutions of the
Euler-Lagrange equations (ELE).

The ${\bf IP}$ with fixed endpoints $t_1,t_2$ was studied first by Weierstrass
who derived a determinant equation \cite{Weier1927}, p. 275, which defines
an existence of conjugate points. Later Howe \cite{Ho1887} applied Weierstrass'
theory to study the PR problem with fixed CL. In the last decades this approach
continued to be used in different setups (see, {\em e.g.}, \cite{Erle1970,
Gil1971}). Stability of axisymmetric menisci with free CL at solid bodies is a
variational ${\bf IP}$ with free endpoints allowed to move along two
given planar curves $S_1,S_2$.

The ${\bf IP}$ providing $\Xi_0[w]\!=\!\min$ with an integrand quadratic in $w,
w',$ linear constraint $\Xi_1[w]\!=\!1$ and fixed BC $w(t_j)\!=\!0$ is related 
to an eigenvalue problem associated with a linear operator (see \cite{CrHl1989},
 Chap. 6). This is true even if the homogeneous (fixed) BC is replaced by any 
other linear homogeneous BC (Dirichlet, Neumann or mixed) and also is consistent
with additional normalization $\Xi_2[w]\!=\!\int_{t_2}^{t_1}\!w^2dt\!=\!1$, 
which eliminates an ambiguity of minimizing solution $Aw(t)$ of ${\bf IP}$ with
arbitrary constant $A$. Thus, the ${\bf IP}$ for the functional $\Xi_0[w]+\mu
\Xi_1[w]-\lambda\Xi_2[w]$ with two Lagrange multipliers $\mu,\lambda$ and two
constraints $\Xi_1,\Xi_2$ gives rise to the Sturm-Liouville equation (SLE) with
real spectrum $\{\lambda_n\}$ and stability criterion: $\min\{\lambda_n\}>0$. A
study of the spectrum $\{\lambda_n\}$ of SLE is a very complicated task for 
generic curves $S_1,S_2$. Such approach was implemented \cite{Myshkis87} to 
study the stability of liquid drop with fixed CL.

Spectral theory of linear operators in PR problem with free CL cannot be applied
directly since a minimization of $\Xi_0[w]$ with constraint $\Xi_1[w]=1$ leads
to a unique solution in the case of inhomogeneous BC $w(t_j)\!\neq 0$. In the
1980s Vogel suggested another approach to this problem constructing an
associated SLE with Neumann BC instead of Dirichlet BC, and established the
stability criterion valid for PR between plates \cite{Vog1987, Vog1989}. This
method requires to solve the eigenvalue problem and to consider the behavior of
the two first minimal eigenvalues $\lambda_{1,2}$. Implementation of this step
is extremely difficult task in the case of ${\sf Und}$ and ${\sf Nod}$ menisci.
This is why only some exact results for ${\sf Cat}$ \cite{Zho97}, ${\sf Sph}$
\cite{Strube91} and ${\sf Und}$ \cite{FinVog1992} menisci between two plates
are known. Investigation of the bridges stability between other surfaces
encounters even more difficulties. This was done only for ${\sf Cyl}$
\cite{Vog1999} and (qualitatively) for convex ${\sf Und}$ and ${\sf Nod}$
\cite{Vog2006} between equal solid spheres. This method \cite{Vog2006} allows to
consider also a stability with respect to non-axisymmetric perturbations. No
results on stability of menisci between other solids (similar or different) are
reported.

Based on the Weierstrass representation of second variation we develop a
non-spectral theory of stability for ${\bf IP}$ with the minimized $E[x,y]$ and
constrained $V[x,y]$ functionals of general type and with free endpoints
belonging to generic curves. We apply this theory to PR of arbitrary shape and
axisymmetric solid bodies to determine the stability of meniscus with free CL.

The present paper is organized in six sections. In the first part, sections
\ref{s2}, \ref{s3}, \ref{s4}, we recall a setup of ${\bf IP}$ and the
Weierstrass representation of second variation with the stability criterion
for variations with fixed endpoints based on which we derive the stability 
criterion for free CL in closed form. In section \ref{s2} we derive two ELE 
supplemented with BC (transversality conditions) and find its extremal solution
 $\bar{x}(t),\bar{y}(t)$ which serves as a functional parameter in formulation 
of ${\bf IP}$ for second variation $\Xi_0[w]\!=\delta^2E[x,y]$ with constraint 
$\Xi_1[w]\!=\delta V[x,y]=0$. In section \ref{s3} for the case of fixed 
endpoints this leads to the Jacobi equation with homogeneous BC $w(t_j)\!=\!0$ 
for perturbation function $w(t)$. Its fundamental and particular solutions 
produce the necessary condition of stability (the criterion of conjugate points 
absence) that generate the stability domain ${\sf Stab}_1(t_2,t_1)$ for extremal
solution in the $\{t_1,t_2\}$-plane. In section \ref{s4} we derive the 
expression for $\delta^2E[x,y]$ as a quadratic form in small perturbations 
$\delta\tau_j$ of the meniscus endpoints along the curves $S_j(\tau_j)$ and 
find a domain ${\mathbb Q}(t_2,t_1)$ where this form is positive definite. 
Finally we find the stability domain ${\sf Stab}_2(t_2,t_1)$ for extremal 
solution with free endpoints as intersection of ${\sf Stab}_1$ and 
${\mathbb Q}$.

In the second part, sections \ref{s5}, \ref{s6}, this approach is applied to
study the stability of axisymmetric PR between solid bodies in absence of
gravity. For ${\sf Cat}$ and ${\sf Cyl}$ menisci we consider different solid
shapes and calculate ${\sf Stab}_2$. Among other new results we verify the
solutions for ${\sf Cat}$ menisci between two plates \cite{Zho97} and ${\sf
Cyl}$ menisci between two spheres \cite{Vog1999} obtained in the framework of
Vogel's theory. The other menisci are treated in section \ref{s62} in a simple
setup between two plates. We find the existence conditions of stable ${\sf Und}$
menisci with and without inflection point and verify conclusions 
\cite{FinVog1992} on their stability for special contact angles. Stability of
${\sf Und}$, ${\sf Nod}$ and ${\sf Sph}$ menisci between non-planar bodies will
be considered in the separate paper.
\section{Stability problem as a variational problem}\label{s2}
Let a planar curve $C$ with parametrization $\{x(t),y(t)\}$, $t_1\leq t\leq
t_2$, be given with its endpoints $\{x(t_j),y(t_j)\}$, $j=1,2$ allowed to move
along two given curves $S_j$ parametrized as $\{X_j(\tau_j),Y_j(\tau_j)\}$, 
$0\leq\tau_j\leq\tau_j^*$ (variable $\tau_j$ runs along $S_j$). Consider the
first isoperimetric problem ({\bf IP-1}) for the functional $E[x,y]$,
\bea
E[x,y]=\int_{t_2}^{t_1}{\sf E}(x,y,x_t,y_t)dt+\sum_{j=1}^2\int_0^{\tau_j^*}
{\sf A}_j(X_j,Y_j,X_{j,\tau},Y_{j,\tau})d\tau,\label{e1}
\eea
with constraint $V[x,y]=1$ imposed on functional,
\bea
V[x,y]=\int_{t_2}^{t_1}{\sf V}(x,y,x_t,y_t)dt+\sum_{j=1}^2(-1)^j\int_0^{
\tau_j^*}{\sf B}_j(X_j,Y_j,X_{j,\tau},Y_{j,\tau})d\tau,\label{e2}
\eea
where we denote $f_t=f'=df/dt$ and $F_{k,t}=F_k'=dF_k/dt$.

The integrands ${\sf E}$ and ${\sf V}$ should be positive homogeneous functions
of degree one in $x_t$ and $y_t$, {\em e.g.}, ${\sf E}(x,y,kx_t,ky_t)=k{\sf E}
(x,y,x_t,y_t)$, that results in identities stemming from Euler theorem for
homogeneous functions,
\bea
{\sf E}=\frac{\partial {\sf E}}{\partial x'}x_t+\frac{\partial {\sf E}}
{\partial y'}y_t,\quad {\sf A}_j=\frac{\partial {\sf A}_j}{\partial X_{j,
\tau_j}}X_{j,\tau_j}+\frac{\partial {\sf A}_j}{\partial Y_{j,\tau_j}}
Y_{j,\tau_j},\label{e3}
\eea
and similar relations for ${\sf V}$ and ${\sf B}_j$.

We have to find an extremal curve $\bar{C}=\{\bar {x}(t),\bar{y}(t)\}$ with
free endpoints $\bar {x}(t_j),\bar{y}(t_j)$ which belong to two given curves 
$S_j$ such that the functional $E[x,y]$ reaches its minimum while the other 
functional $V[x,y]$ is constrained.

Define a functional $W[x,y]=E[x,y]-\lambda V[x,y]$ with the Lagrange multiplier
$\lambda$
\bea
W[x,y]=\int_{t_2}^{t_1}F(x,y,x_t,y_t)dt-\sum_{j=1}^2(-1)^j\int_0^{\tau_j^*}
G_j(X_j,Y_j,X_{j,\tau},Y_{j,\tau})d\tau,\label{e4}
\eea
where $F={\sf E}-\lambda{\sf V}$, $G_1=\lambda{\sf B}_1+{\sf A}_1$, $G_2=
\lambda{\sf B}_2-{\sf A}_2$. According to (\ref{e3}) we have also
\bea
F=\frac{\partial F}{\partial x_t}x_t+\frac{\partial F}{\partial y_t}y_t,\quad
G_j=\frac{\partial G_j}{\partial X_{j,\tau_j}}X_{j,\tau_j}+\frac{\partial G_j}
{\partial Y_{j,\tau_j}}Y_{j,\tau_j}.\label{e5}
\eea
Calculate the total variation of the functional, ${\mathbb D}W={\mathbb D}_0W+
{\mathbb D}_1W-{\mathbb D}_2W$,
\bea
{\mathbb D}_0W&=&\int_{t_2+\delta t_2}^{t_1+\delta t_1}[F+\Delta_1 F+\Delta_2 F+
\ldots ]\;dt-\int_{t_2}^{t_1}Fdt,\label{e6}\\
{\mathbb D}_jW&=&\int_0^{\tau_j^*+\delta\tau_j}G_jd\tau_j-\int_0^{\tau_j^*}G_jd
\tau_j,\nonumber
\eea
where
\bea
\Delta_1F&=&\frac{\partial F}{\partial x}u+\frac{\partial F}{\partial x_t}u'+
\frac{\partial F}{\partial y}v+\frac{\partial F}{\partial y_t}v',\label{e7}\\
\Delta_2F&=&\frac{u^2}{2}\frac{\partial^2F}{\partial x^2}+uu'\frac{\partial^2 F}
{\partial x\partial x_t}+\frac{u'^2}{2}\frac{\partial^2 F}{\partial x_t^2}+
\frac{v^2}{2}\frac{\partial^2 F}{\partial y^2}+vv'\frac{\partial^2 F}
{\partial y\partial y_t}\nonumber\\
&+&\frac{v'^2}{2}\frac{\partial^2 F}{\partial y_t^2}+uv\frac{\partial^2 F}
{\partial x\partial y}+uv'\frac{\partial^2 F}{\partial x\partial y_t}+u'v\frac{
\partial^2F}{\partial x_t\partial y}+u'v'\frac{\partial^2F}{\partial x_t
\partial y_t},\nonumber
\eea
and $\{u(t),v(t)\}$ is a small perturbation in vicinity of curve $\bar{C}$
where the extremum of {\bf IP-1} is reached. Define a projection of the $\{u(t),
v(t)\}$ vector on the normal to the extremal $\{\bar{x}(t),\bar{y}(t)\}$,
\be
w(t)=u\;\bar{y}_t-v\;\bar{x}_t.\label{e8}
\ee
Represent ${\mathbb D}_0W$ and ${\mathbb D}_jW$ up to the terms quadratic in
$\delta\tau_j,u,v,u',v'$,
\bea
&&{\mathbb D}_0W=\int_{t_2}^{t_1}\!\!\Delta_1 Fdt+\int_{t_2}^{t_1}\!\!\Delta_2
Fdt,\quad {\mathbb D}_jW=G_j^*\delta\tau_j+\frac1{2}\frac{dG_j^*}{d\tau_j}
\delta^2\tau_j,\label{e9}\\
&&\frac{dG_j^*}{d\tau_j}=\frac{\partial G_j^*}{\partial X_j}\frac{dX_j}{d\tau_j}
+\frac{\partial G_j^*}{\partial Y_j}\frac{dY_j}{d\tau_j}+\frac{\partial G_j^*}
{\partial X_j'}\frac{d^2X_j}{d\tau_j^2}+\frac{\partial G_j^*}{\partial Y_j'}
\frac{d^2Y_j}{d\tau_j^2},\nonumber
\eea
where $G_j^*=G_j$ and $\partial G_j^*/\partial X_j=\partial G_j/\partial X_j$ 
computed at $\tau_j=\tau_j^*$.
\subsection{First variation $\delta W$ and ELE}\label{s21}
Using the terms in (\ref{e9}) linear in $\delta\tau_j$, $u,v$ and $u_t,v_t$,
in expression (\ref{e6}) calculate $\delta W$
\be
\delta W=\int_{t_2}^{t_1}\!\!\!\Delta_1 Fdt+G_1^*\delta\tau_1-G_2^*\delta\tau_2.
\label{e10}
\ee
To derive BC for perturbations $u(t_j),v(t_j)$ we have to make them consistent
with free endpoints running along the curves $S_j$
\bea
&&\bar{x}(t_j)=X(\tau_j^*),\quad\bar{x}(t_j)+u(t_j)=X(\tau_j^*+\delta\tau_j),
\label{e11}\\
&&\bar{y}(t_j)\!=\!Y(\tau_j^*),\hspace{.5cm}\bar{y}(t_j)+v(t_j)=Y(\tau_j^*+
\delta\tau_j),\nonumber
\eea
resulting in a sequence of equalities: $u(t_j)=\sum_{k=1}^{\infty}u_k(t_j)$ and
$v(t_j)=\sum_{k=1}^{\infty}v_k(t_j)$,
\bea
u_k(t_j)=\frac1{k!}\frac{d^kX_j}{d\tau_j^k}\delta^k\tau_j,\quad v_k(t_j)=
\frac1{k!}\frac{d^kY_j}{d\tau_j^k}\delta^k\tau_j.\label{e12}
\eea
The function $w(t)$ defined in (\ref{e8}) reads at the endpoints,
\bea
w(t_j)=\eta(t_j,\tau_j^*)\delta\tau_j,\quad\eta(t_j,\tau_j^*)=\bar{y}_t\frac{
dX_j}{d\tau_j}-\bar{x}_t\frac{dY_j}{d\tau_j}.\label{e13}
\eea
Denote by $\delta F/\delta z=\partial F/\partial z-\frac{d}{dt}(\partial F/
\partial z')$ the variational derivative. Then $\delta W$ in (\ref{e10}) may be
written as
\bea
\delta W=\int_{t_2}^{t_1}\left(u\frac{\delta F}{\delta x}+v\frac{\delta F}
{\delta y}\right)dt+\left[u_1\frac{\partial F}{\partial x'}+v_1\frac{\partial F}
{\partial y'}\right]_{t_2}^{t_1}+G_1^*\delta\tau_1-G_2^*\delta\tau_2.
\nonumber
\eea
Substitute $u_1(t_j)$ and $v_1(t_j)$ from (\ref{e12}) into the last expression
and obtian
\bea
\delta W=\int_{t_2}^{t_1}\!\left(u\frac{\delta F}{\delta x}+v\frac{\delta F}
{\delta y}\right)dt-\sum_{j=1}^2(-1)^j\left[\frac{\partial F_j}{\partial x'}
\frac{dX_j}{d\tau_j}+\frac{\partial F_j}{\partial y'}\frac{dY_j}{d\tau_j}+G_j^*
\right]\delta\tau_j,\nonumber
\eea
where $F_j=F,\;\partial F_j/\partial x = \partial F/\partial x,$ {\it etc.}
computed at $t=t_j$. Thus, we arrive at ELE
\bea
\frac{\partial F}{\partial x}-\frac{d}{dt}\frac{\partial F}{\partial x'}=0,\quad
\frac{\partial F}{\partial y}-\frac{d}{dt}\frac{\partial F}{\partial y'}=0,
\label{e14}
\eea
supplemented by the transversality conditions:
\bea
\frac{\partial F_2}{\partial x'}\frac{dX_2}{d\tau_2}+\frac{\partial F_2}
{\partial y'}\frac{dY_2}{d\tau_2}+G_2^*=0,\quad
\frac{\partial F_1}{\partial x'}\frac{dX_1}{d\tau_1}+\frac{\partial F_1}
{\partial y'}\frac{dY_1}{d\tau_1}+G_1^*=0.\label{e15}
\eea
Solution $\bar{x}(t),\bar{y}(t)$ provides the extremal value of $E[x,y]$ and 
constraint $V[x,y]=1$.

Identify $E[x,y]$ as a functional of surface energy of PR and fix its volume by
variational constraint $V[x,y]=1$. Then we arrive at the PR problem \cite{RF13}
in absence of gravity where ELE (\ref{e14}) and transversality conditions
(\ref{e15}) are known as YLE and Young relations. The latter leaves free the
values $x(t_j)$, $y(t_j)$ at the endpoints where the meniscus contacts the
solid surfaces at the fixed contact angles.
\subsection{The Weierstrass representation of second variation $\delta^2W$}
\label{s22}
Making use in (\ref{e6}) of the terms quadratic in $\delta\tau_j$, $u,v$ and
$u',v'$, calculate the second variation $\delta^2W$,
\bea
\delta^2W=\int_{t_2}^{t_1}\Delta_2 Fdt+\left(\frac{\partial F}{\partial x'}
u_2+\frac{\partial F}{\partial y'}v_2\right)_{t_2}^{t_1}+\frac1{2}\left(
\frac{dG_1}{d\tau_1}\delta^2\tau_1-\frac{dG_2}{d\tau_2}\delta^2\tau_2\right),
\label{e16}
\eea
Substituting $u_2$ and $v_2$ from (\ref{e12}) into the last expression we obtain
\bea
\delta^2W=\int_{t_2}^{t_1}\Delta_2 Fdt-\frac1{2}\sum_{j=1}^2(-1)^j\left(
\frac{\partial F}{\partial x'}\frac{d^2X_j}{d\tau_j^2}+\frac{\partial F}
{\partial y'}\frac{d^2Y_j}{d\tau_j^2}+\frac{dG_j}{d\tau_j}\right)\delta^2
\tau_j.\nonumber
\eea
Denote $\delta^2_BW=\int_{t_2}^{t_1}\Delta_2 Fdt$ and following Weierstrass
\cite{Weier1927}, pp.132-134 (see also Bolza \cite{Bolz1904}, p.206) represent
$\delta^2_BW$ in terms of small perturbation $\{u(t),v(t)\}$ of the extremal
curve $\{\bar{x}(t),\bar{y}(t)\}$ and $w(t)$,
\bea
&&\delta^2_BW[x,y]=\frac1{2}\Xi_0[w]+\frac1{2}\left[Lu_1^2+2Mu_1v_1
+Nv_1^2\right]|_{t_2}^{t_1},\label{e17}\\
&&\Xi_0[w]\!=\!\int_{t_2}^{t_1}\!\!\left[H_1w'^2\!+\!H_2w^2\right]dt,\;M\!=\!
F_{xy'}+\bar{x}_t\bar{y}_{tt}H_1\!=\!F_{yx'}+\bar{y}_t\bar{x}_{tt}H_1,
\hspace{.5cm}\label{e18}\\
&&L\!=\!F_{xx'}-\bar{y}_t\bar{y}_{tt}H_1,\;N\!=\!F_{yy'}-\bar{x}_t\bar{x}_{tt}
H_1,\;H_1\!=\!\frac{F_{x'x'}}{\bar{y}_{t}^2}\!=\!\frac{F_{y'y'}}{\bar{x}_t^2}
\!=\!-\frac{F_{x'y'}}{\bar{x}_t\bar{y}_{t}},\nonumber\\
&&H_2=\frac{F_{xx}-\bar{y}_{tt}^2H_1-L_t}{\bar{y}_t^2}=\frac{F_{yy}-
\bar{x}_{tt}^2H_1-N_t}{\bar{x}_t^2}=-\frac{F_{xy}+\bar{x}_{tt}\bar{y}_{tt}
H_1-M_t}{\bar{x}_t\bar{y}_{t}}.\nonumber
\eea
Substituting (\ref{e12}) and (\ref{e9}) into (\ref{e16}) we obtain,
\bea
&&\delta^2W=\delta^2_BW+\xi_1\delta^2\tau_1-\xi_2\delta^2\tau_2,\quad
\mbox{where}\label{e19}
\eea
\bea
2\xi_j=\frac{\partial F_j}{\partial x'}\frac{d^2X_j}{d\tau_j^2}+
\frac{\partial F_j}{\partial y'}\frac{d^2Y_j}{d\tau_j^2}+
\frac{\partial G_j}{\partial X_j}\frac{dX_j}{d\tau_j}+
\frac{\partial G_j}{\partial Y_j}\frac{dY_j}{d\tau_j}+
\frac{\partial G_j}{\partial X_j'}\frac{d^2X_j}{d\tau_j^2}+
\frac{\partial G_j}{\partial Y_j'}\frac{d^2Y_j}{d\tau_j^2}.\nonumber
\eea
Substitute $u_1(t_j),v_1(t_j)$ from (\ref{e12}) into (\ref{e17}) and combine
it with (\ref{e19}), and find,
\bea
&&\delta^2W=\frac1{2}\Xi_0[w]+K_1\delta^2\tau_1-K_2\delta^2\tau_2,\label{e20}\\
&&
2K_j=2\xi_j+L(t_j)\left(\frac{dX_j}{d\tau_j}\right)^2+2M(t_j)\frac{dX_j}
{d\tau_j}\frac{dY_j}{d\tau_j}+N(t_j)\left(\frac{dY_j}{d\tau_j}\right)^2\!.
\quad\label{e21}
\eea
\section{Homogeneous boundary conditions: fixed endpoints}\label{s3}
Study the stability of the extremal curve $\{\bar{x}(t),\bar{y}(t)\}$ w.r.t.
small fluctuations in two different cases considered separately; the first case 
corresponds to the perturbation of the extremal curve in the interval $(t_2,
t_1)$ for the fixed endpoints,
\bea
u(t_j)=v(t_j)=w(t_j)=0,\quad j=1,2.\label{f1}
\eea
The second case is when at least one endpoint is free and allowed to run along
given curves $S_j$ is discussed in section \ref{s4}. Start with the second
isoperimetric problem ({\bf IP-2}) associated with perturbations $\{u(t),v(t)\}$
in the vicinity of $\{\bar{x}(t),\bar{y}(t)\}$ with BC (\ref{f1}). Following
Bolza \cite{Bolz1904}, p.215, write the constraint for $V[x,y]$,
\bea
\Xi_1[w]=\int_{t_2}^{t_1}\!\!H_3wdt=0,\quad\left\{\begin{array}{l}H_3={\sf V}_
{xy'}-{\sf V}_{x'y}+H_4(\bar{x}_t\bar{y}_{tt}-\bar{y}_{t}\bar{x}_{tt}),\\
H_4={\sf V}_{x'x'}\bar{y}_{t}^{-2}={\sf V}_{y'y'}\bar{x}_t^{-2}=-{\sf V}_{x'y'}
\bar{x}_t^{-1}\bar{y}_t^{-1},\end{array}\right.\label{f2}
\eea
which involves perturbation $w$. For PR problem we have ${\sf V}=x^2y',\;
{\sf B}_j=X_j^2Y_j'$, leading to $H_3=\bar{x}$, which substantially simplifies
the computation (see section \ref{s5}). 

Substitute (\ref{f1}) into (\ref{e17}) and arrive at the classical ${\bf IP}$ 
with the second variation $\Xi_0[w]$ treated in the framework of Weierstrass' 
theory (see \cite{Bolz1904}, Chap. 6). Analyzing the problem with functional 
$\Xi_2[w]=\Xi_0[w]+2\mu\Xi_1[w]$,
\bea
\Xi_2[w]=\int_{t_2}^{t_1}{\mathcal H}(t,w,w')dt,\quad {\mathcal H}(t,w,w')=
H_1w'^2+H_2w^2+2\mu H_3w,\label{f3}
\eea
and the Lagrange multiplier $\mu$, write ELE for the function $w(t)$ as an
inhomogeneous Jacobi equation with BC given in (\ref{f1})
\bea
(H_1w')'-H_2w=\mu H_3,\quad w(t_1)=w(t_2)=0.\label{f4}
\eea
The point $t_2'\neq t_2$ is called {\it conjugate} to the point $t_2,$ if
(\ref{f4}) has a solution $\bar{w}(t)$ such that $\bar{w}(t_2)=\bar{w}(t_2')=
0,$ but is not identically zero. According to Bolza \cite{Bolz1904}, pp.217-220,
the following set of conditions is sufficient for the functional (\ref{f3}) to 
have a weak minimum for the solution $\bar{w}(t)$ of equations (\ref{f4}):
\bea
a)\;\;H_1(t)>0\;,\quad b)\;\;\;\mbox{the interval}\;\;[t_2,t_1]\;\mbox{
contains no points conjugate to}\;t_2\;.\label{f5}
\eea
In fact, conditions (\ref{f5}) provide a strong minimum because the Weierstrass
function ${\mathcal E}(t,w,w',f)={\mathcal H}(t,w,f)-{\mathcal H}(t,w,w')+(w'-f)
{\mathcal H}_{w'}(t,w,w')$ for the functional $\Xi_2[w]$ in (\ref{f3}) is 
positive,
\bea
{\mathcal E}(t,w,w',f)=H_1\left[f(t)-w'(t)\right]^2,\quad\mbox{for}\quad f(t)
\neq w'(t).\label{f6}
\eea

Weierstrass \cite{Weier1927}, p. 275, gave another version of conjugate points
non-existence condition. Assume that $\bar{w}_1(t)$ and $\bar{w}_2(t)$ are
fundamental solutions of homogeneous Jacobi equation, then the particular
solution $\mu\bar{w}_3(t)$ of inhomogeneous Jacobi equation (\ref{f4}) may be
found by standard procedure
\be
\bar{w}_3(t)=\bar{w}_2\int^t\frac{\bar{w}_1H_3}{H_1{\sf Wr}}\;ds-\bar{w}_1
\int^t\frac{\bar{w}_2H_3}{H_1{\sf Wr}}\;ds,\quad {\sf Wr}=\bar{w}_1\bar{w}_2'-
\bar{w}_2\bar{w}_1',\label{f7}
\ee
where ${\sf Wr}$ denotes the Wronskian for fundamental solutions. Find ${\sf
Wr}$ assuming that $\bar{w}_1$ is known and the second fundamental solution
reads $\bar{w}_2=U(t)\bar{w}_1$. Substitute it into (\ref{f4}) with $\mu=0$ and
obtain
\bea
\frac{d}{dt}\left(H_1\bar{w}_1^2\frac{dU}{dt}\right)=0,\quad\frac{dU}{dt}=
\frac{g}{H_1\bar{w}_1^2},\quad{\sf Wr}=\bar{w}_1^2\frac{dU}{dt}=\frac{g}{H_1},
\label{f16}
\eea
where $g$ is an integration constant. The fundamental solutions $w_j$ also can
be expressed as $w_j=y'(\partial x/\partial \alpha_j)-x'(\partial y/\partial
\alpha_j),\;j=1,2,$ where $\alpha_j$ denotes the integration constant emerging
from ELE (see Bolza \cite{Bolz1904}, p.219). Making use of the last expression
in (\ref{f7}) we arrive at
\be
g\bar{w}_3=\bar{w}_2J_1-\bar{w}_1J_2,\quad g\bar{w}_3'=\bar{w}_2'J_1-\bar{w}_1'
J_2,\quad g\bar{w}_3''=\bar{w}_2''J_1-\bar{w}_1''J_2+\frac{gH_3}{H_1},
\label{f17}
\ee
where $J_k=\int^tH_3\bar{w}_k ds$. Following Weierstrass \cite{Weier1927}
introduce the matrix,
\bea
D(t_2,t')\!=\!\left(\begin{array}{ccc}\bar{w}_1(t_2)&\bar{w}_2(t_2)&\bar{w}_3
(t_2)\!\\\bar{w}_1(t') & \bar{w}_2(t')& \bar{w}_3(t')\!\\
J_1(t')-J_1(t_2) & J_2(t')-J_2(t_2) & J_3(t')-J_3(t_2)\!\end{array}\right).
\label{f8}
\eea
Then the condition of non-existence of conjugate points reads (see 
\cite{Weier1927}, p.275),
\bea
\Delta(t_2,t')\neq 0,\quad t_2<t'<t_1,\quad\Delta(t_2,t_1)=\det D(t_2,t_1).
\label{f9}
\eea
Bolza in \cite{Bolz1904}, p.223, gave a more general condition of non-existence
of conjugate points,
\bea
\Delta(t'',t')\neq 0,\quad t_2<t'<t''<t_1,\label{f10}
\eea
making the Jacobi condition (\ref{f5}b) symmetric with respect to the endpoints
$t_2$ and $t_1$. Write a determinant equation $\Delta(t_2,t_1)=0$ as follows,
\bea
\Delta(t_2,t_1)=I_3\left[\bar{w}_1(t_2)\bar{w}_2(t_1)-\bar{w}_1(t_1)\bar{w}_2
(t_2)\right]+\left[I_1\bar{w}_2(t_1)-I_2\bar{w}_1(t_1)\right]\bar{w}_3(t_2)
\label{f11}\\
+\left[I_2\bar{w}_1(t_2)-I_1\bar{w}_2(t_2)\right]\bar{w}_3(t_1),\quad
\mbox{where}\quad I_k=J_k(t_1)-J_k(t_2).\nonumber
\eea
If $\bar{w}_1(t)$ and $\bar{w}_2(t)$ are continuous functions then solution of 
equation $\Delta(t_2,t_1)=0$ describes a continuous curve ${\cal D}(t_2,t_1)$
of conjugated points.

Another important requirement is to guarantee that the extremal $\{\bar{x}(t),
\bar{y}(t)\}$ does not intersect with the curves $S_j$. In the case of the PR 
this requirement provides the meniscus existence condition given by the constant
sign of $\eta(t_j,\tau_j^*)$. Define the lines $t_j=t_j^{\bullet}$ in $\{t_1,
t_2\}$-plane where $\eta(t_j^{\bullet},\tau_j^*)=0$.

Consider a point $M_1=(a,b)$ in the lower halfplane $\{t_2<t_1\}$ and two
more points: $M_2=(a,a)$ and $M_3=(b,b)$. Call a point $M_1$ the Jacobi
point if the line $M_1M_2$ does not intersect both ${\cal D}(t_2,t_1)$ and
$t_2=t_2^{\bullet}$, and $M_1M_3$ does not intersect both ${\cal D}(t_2,t_1)$
and $t_1=t_1^{\bullet}$. Define a set ${\mathbb J}(t_2,t_1)$ as a union
of points $M_1$
\bea
{\mathbb J}(t_2,t_1)=\left\{(a,b)\;\left\bracevert\left.\begin{array}{l}
\;\Delta(t,a)\neq 0,\;\Delta(b,t)\neq 0,\;\;t_2<b\leq t\leq a<t_1,\\
\eta(t_2,\tau_2^*)\neq 0,\;\;t_2^{\bullet}<b\leq t_2\leq a<t_1,\\
\eta(t_1,\tau_1^*)\neq 0,\;\;t_2<b\leq t_1\leq a<t_1^{\bullet}.\end{array}
\right.\right.\right\}\label{f12}
\eea
representing an open domain in $\{t_1,t_2\}$-plane. Combining (\ref{f5} a,b) and
(\ref{f12}) define a stability set as intersection set
\bea
{\sf Stab}_1(t_2,\!t_1)\!=\!{\mathbb J}(t_2,\!t_1)\cap {\mathbb L}(t_2,\!t_1),\;
\;{\mathbb L}(t_2,\!t_1)\!=\!\left\{(t_2,t_1)|H_1(t)>0,\;t\in[t_2,\!t_1]\right\}
\label{f13}
\eea
where the set ${\mathbb L}(t_2,t_1)$ comprises the points satisfying Legendre's
criterion (\ref{f5}a).
\section{Inhomogeneous boundary conditions: free endpoints}\label{s4}
Consider the case when the extremal $\{\bar{x}(t),\bar{y}(t)\}$ is perturbed at
the interval $[t_2,t_1]$ including both endpoints. The case of one free and one
fixed endpoints will follow as a corollary. The nonintegral term in (\ref{e17})
is fixed and in general case it does not vanish; the same is true 
for (\ref{e20}). 
It is worth to mention that any other BC, {\em e.g.}, the Neumann BC $w'(t_j)
=0$ in \cite{Vog1987} or mixed BC $g_1w'(t_j)+g_0w(t_j)=0$ in \cite{Myshkis87},
leads to changes in $u(t_j)$, $v(t_j)$ and requires variation of the nonintegral
term in (\ref{e17}).

From physical point of view BC (\ref{e13}) requires that {\em the endpoints of
perturbed meniscus $\{\bar{x}+u,\bar{y}+v\}$ always belong to the solid
surfaces}. These claims are justified from mathematical standpoint:
\begin{itemize}
\item The second order Jacobi equation (\ref{f4}) for perturbation $w$ admits
no more than two BC.
\item The perturbed meniscus $\{\bar{x}+u,\bar{y}+v\}$ may not provide the
extremum for $W[x,y]$ even if $\{u,v\}$ do provide the extremum for
$\delta^2W[x,y]$.
\end{itemize}

Following an ideology of stability theory we have to find when $\delta^2W$ is
positive definite in vicinity of the extremal curve constrained by (\ref{e2}).
Since the only varying part in (\ref{e20}) is the functional $\Xi_0[w]$, this
brings us to {\bf IP-2} with one indeterminate function $w(t)$: find the
extremal $\bar{w}(t)$ providing $\Xi_0[w]$ to be positive definite in vicinity
of $\bar{w}(t)$ and preserving $\Xi_1[w]$. Inhomogeneity of BC requires to
answer two questions:
\bea
&&\hspace{-2cm}
\mbox{{\sl When is $\Xi_0[w]$ positive definite in vicinity of $\bar{w}(t)$
for the fixed $\delta\tau_j$ ?}}\label{g1}\\
&&\hspace{-2cm}
\mbox{{\sl When does $\Xi_0[\bar{w}]$ reach a positive value as a function of
displacements $\delta\tau_j$ ?}}\label{g2}
\eea

Start with (\ref{g1}) and consider the necessary conditions for functional
$\Xi_0[w]$ to be positive definite in vicinity of extremal perturbation $\bar{w}
(t)$ for the fixed $\delta\tau_j$ and preserving $\Xi_1[w]$. Let us prove that 
they coincide with those conditions (\ref{f5}) for the functional $E[x,y]$ to be
positive definite in vicinity of extremal solution $\{\bar{x}(t),\bar{y}(t)\}$
for the fixed endpoints and preserving $V[x,y]$.

For this purpose we ignore a fact that $\Xi_0[w]$ is a second variation,
satisfying the relations (\ref{e18}), and instead, we treat the analysis of
(\ref{e18}) as independent problem. Represent $w$ in a vicinity of extremal
perturbation $\bar{w}$,
\bea
w(t)=\bar{w}(t)+\varepsilon(t),\quad\varepsilon(t_1)=\varepsilon(t_2)=0,\quad
\Xi_1[\varepsilon]=\int_{t_2}^{t_1}H_3\varepsilon dt=0,\label{g3}
\eea
where a perturbation $\varepsilon$ preserves both BC (\ref{e13}) and the
constraint (\ref{f2}). Find the first and second variations of functional
$\Xi_2[w]$ defined in (\ref{f3}),
\bea
\delta\Xi_2[w]\!=\!2\int_{t_2}^{t_1}\left[-\left(H_1\bar{w}'\right)'+H_2\bar{w}
+\mu H_3\right]\varepsilon\;dt,\quad\delta^2\Xi_2[w]\!=\!\int_{t_2}^{t_1}
\left[H_1\varepsilon'^2+H_2\varepsilon^2\right]dt.\nonumber
\eea
The first variation $\delta\Xi_2[w]$ vanishes at the extremal $\bar{w}$
satisfying the inhomogeneous Jacobi equation (\ref{f4}). Regarding the second
variation $\delta^2\Xi_2[w]$, it completely coincides with $\Xi_0[w]$ as well as
BC and volume constraint (\ref{g3}) are coinciding with similar BC (\ref{f1})
and constraint (\ref{f2}) in the {\bf IP} with fixed endpoints (section 
\ref{s3}). This coincidence implies the necessary conditions (\ref{f5}) for 
$\Xi_0[w]$ to be positive definite in vicinity of extremal $\bar{w}$ for the 
fixed $\delta\tau_j$.

Consider (\ref{g2}) and write a general solution $\bar{w}$ of equation
(\ref{f4}) built upon the fundamental solutions $\bar{w}_1,\bar{w}_2$ of
homogeneous equation, and particular solution of inhomogeneous equation $\bar{
w}_3$,
\be
\bar{w}(t)=C_1\bar{w}_1(t)+C_2\bar{w}_2(t)+\mu\bar{w}_3(t)\;.\label{g6}
\ee
Inserting (\ref{g6}) into BC (\ref{e13}) and into constraint (\ref{f2}) we
obtain three linear equations,
\be
\bar{w}_1(t_j)C_1+\bar{w}_2(t_j)C_2+\bar{w}_3(t_j)\mu=\bar{w}(t_j),\quad
I_1C_1+I_2C_2+I_3\mu=0,\label{g7}
\ee
which are uniquely solvable (see \cite{Bolz1904}, p.220) if $\Delta(t_2,t_1)\neq
0$ and have nonzero solutions when at least one of $\bar{w}(t_j)$ is nonzero,
\bea
C_j=m_{j1}\delta\tau_1+m_{j2}\delta\tau_2,\;j=1,2,\quad\mu=m_{31}\delta\tau_1
+m_{32}\delta\tau_2.\label{g8}
\eea
Substitute (\ref{g8}) into (\ref{g7}) and find two equations with matrix
$D(t_2,t_1)$ defined in (\ref{f8}),
\bea
D(t_2,t_1){\bf M}_j={\bf N}_j,\quad {\bf M}_j=\left(\begin{array}{c}m_{1j}\\
m_{2j}\\m_{3j}\end{array}\right),\quad {\bf N}_1=\left(\begin{array}{c}0\\
\eta_1\\0\end{array}\right),\quad {\bf N}_2=\left(\begin{array}{c}\eta_2\\0\\
0\end{array}\right),\nonumber
\eea
where $\eta_j=\eta(t_j,\tau_j^*)$ and $\bar{w}_i(t_j)=\bar{w}_{ij}$. Then
$m_{j1}=\eta_1\beta_{j1}/\Delta$, $m_{j2}=\eta_2\beta_{j2}/\Delta,$ and
\bea
\beta_{11}=I_3\bar{w}_{22}-I_2\bar{w}_{32},\quad
\beta_{21}=I_1\bar{w}_{32}-I_3\bar{w}_{12},\quad
\beta_{31}=I_2\bar{w}_{12}-I_1\bar{w}_{22},\nonumber\\
\beta_{12}=I_2\bar{w}_{31}-I_3\bar{w}_{21},\quad
\beta_{22}=I_3\bar{w}_{11}-I_1\bar{w}_{31},\quad
\beta_{32}=I_1\bar{w}_{21}-I_2\bar{w}_{11}.\nonumber
\eea
Substituting (\ref{g8}) into (\ref{g6}), represent $\bar{w}(t)$ as follows
\bea
\bar{w}(t)=A_1(t)\delta\tau_1+A_2(t)\delta\tau_2,\quad A_j(t)=\frac{\eta_j
B_j(t)}{\Delta(t_2,t_1)},\quad B_j(t)=B_j(t,t_2,t_1),\label{g11}
\eea
\bea
B_1(t)=-\left|\begin{array}{ccc}\bar{w}_1(t)&\bar{w}_2(t)&\bar{w}_3(t)\\
\bar{w}_1(t_2) & \bar{w}_2(t_2)& \bar{w}_3(t_2)\\
I_1 & I_2 & I_3\end{array}\right|,\quad
B_2(t)\!=\!\left|\begin{array}{ccc}\bar{w}_1(t)&\bar{w}_2(t)&\bar{w}_3(t)\\
\bar{w}_1(t_1) & \bar{w}_2(t_1)& \bar{w}_3(t_1)\\
I_1 & I_2 & I_3\end{array}\right|.\nonumber
\eea
According to (\ref{e13}) we have, $B_1(t_2)=B_2(t_1)=0$, $B_j(t_j)=\Delta(t_2,
t_1)$, and its expression is given in (\ref{f11}). Straightforward calculation
of determinant's derivatives gives
\bea
&&H_1(t_2)B_1'(t_2)=-H_1(t_1)B_2'(t_1)=I_1(t_1)I_2(t_2)-I_1(t_2)I_2(t_1)+gI_3,
\nonumber\\
&&gB_1'(t_1)=\left[I_2\bar{w}_1(t_2)-I_1\bar{w}_2(t_2)\right]
\left[I_2\bar{w}_1'(t_1)-I_1\bar{w}_2'(t_1)\right]\nonumber\\
&&\hspace{3cm}-H_1(t_1)B_2'(t_1)[\bar{w}_1(t_2)\bar{w}_2'(t_1)-
\bar{w}_2(t_2)\bar{w}_1'(t_1)],\nonumber\\
&&gB_2'(t_2)=\left[I_2\bar{w}_1(t_1)-I_1\bar{w}_2(t_1)\right]
\left[I_2\bar{w}_1'(t_2)-I_1\bar{w}_2'(t_2)\right]\nonumber\\
&&\hspace{3cm}-H_1(t_2)B_1'(t_2)[\bar{w}_1(t_1)\bar{w}_2'(t_2)-
\bar{w}_2(t_1)\bar{w}_1'(t_2)],\label{g12d}
\eea
where $B_j'(t_k)\equiv dB_j(t,t_2,t_1)/dt$ computed at $t=t_k$. Formula
(\ref{e17}) together with equation (\ref{f4}) allows to express $\delta^2W[x,y]$ in
a simple form. Multiplying (\ref{f4}) by $\bar{w}$ and integrating by parts we
obtain
$$
\int_{t_2}^{t_1}\left[H_1(t)\bar{w}'^2(t)+H_2(t)\bar{w}^2(t)\right]dt-
H_1(t)\bar{w}(t)\bar{w}'(t)|_{t_2}^{t_1}=0.
$$
Combining the last equality with (\ref{e20}) and (\ref{e21}) we arrive at
\be
\delta^2W=\frac1{2}\left[H_1\bar{w}\bar{w}'+Lu^2+2Muv+Nv^2\right]|_{t_2}^{t_1}
+\xi_1\delta^2\tau_1-\xi_2\delta^2\tau_2,\label{g13}
\ee
where $\xi_j$ are defined in (\ref{e19}). Substituting (\ref{e12}, \ref{g11})
into (\ref{g13}) and using (\ref{g12d}), we obtain
\bea
&&\delta^2W=Q_{11}\left(\delta\tau_1\right)^2+2Q_{12}\delta\tau_1\delta\tau_2+
Q_{22}\left(\delta\tau_2\right)^2,\label{g14}\\
&&Q_{11}(t_2,t_1)=\frac{\eta_1^2P_{11}}{2\Delta}+K_1,\hspace{1cm}
P_{11}=H_1(t_1)B_1'(t_1),\nonumber\\
&&Q_{22}(t_2,t_1)=\frac{\eta_2^2P_{22}}{2\Delta}-K_2,\hspace{1cm}
P_{22}=-H_1(t_2)B_2'(t_2),\nonumber\\
&&Q_{12}(t_2,t_1)=\frac{\eta_1\eta_2P_{12}}{2\Delta},\hspace{1.6cm}
P_{12}=P_{21}=H_1(t_1)B_2'(t_1),\label{g15}
\eea
where $\eta_j=\eta(t_j,\tau_j^*)$ and $K_j=K_j(t_j,\tau_j^*)$ are defined in
(\ref{e13}) and (\ref{e21}), respectively.

Using BC (\ref{e11}): $\bar{x}(t_j)=X(\tau_j^*)$, $\bar{y}(t_j)=Y(\tau_j^*)$,
the matrix elements $Q_{ij}$ may be represented as functions of $t_2,t_1$ only.
The necessary conditions to have $\delta^2W\geq 0$ are given by three
inequalities,
\bea
Q_{11}(t_2,t_1)\geq 0,\quad Q_{22}(t_2,t_1)\geq 0,\quad Q_{33}(t_2,t_1)=
Q_{11}Q_{22}-Q_{12}^2\geq 0.\label{g16}
\eea
One of the two first inequalities in (\ref{g16}) is redundant but we leave it
for the symmetry considerations. Inequalities (\ref{g16}) provide an answer to
the question (\ref{g2}). Define three different sets ${\mathbb Q}_j(t_2,t_1)$
\bea
{\mathbb Q}_j(t_2,t_1):=\left\{(a,b)\;|\;(a,b)\in\{t_2<t_1\},\;Q_{jj}(t_2,t_1)
\geq 0\right\},\label{g17}
\eea
and the intersection set ${\mathbb Q}(t_2,t_1):={\mathbb Q}_1(t_2,t_1)\cap
{\mathbb Q}_2(t_2,t_1)\cap{\mathbb Q}_3(t_2,t_1)$.

Summarizing answers to both questions (\ref{g1}, \ref{g2}) we conclude that the
necessary conditions of stability of extremal $\bar{w}(t)$ with BC comprise
(\ref{f5}), (\ref{f9}), (\ref{f13}) and (\ref{g15}):
\bea
{\sf Stab}_2(t_2,t_1)={\sf Stab}_1(t_2,t_1)\cap{\mathbb Q}(t_2,t_1),\quad
{\sf Stab}_2(t_2,t_1)\subseteq {\sf Stab}_1(t_2,t_1).\label{g18}
\eea
The conditions (\ref{g16}) cannot determine the extremal
solution stability in case when the determinant $Q_{33}$ in (\ref{g16}) vanishes.
Indeed, we have in (\ref{g14})
\bea
\delta^2W=\left(\sqrt{Q_{11}}\delta\tau_1+\sqrt{Q_{22}}\delta\tau_2\right)^2,
\quad Q_{33}(t_2,t_1)=0.\label{g19}
\eea
Thus, there exists a non empty set of perturbations $\delta\tau_1,\delta\tau_2$
such that $\sqrt{Q_{11}}\delta\tau_1+\sqrt{Q_{22}}\delta\tau_2=0,$ which does not
affect the second variation, {\em i.e.}, $\delta^2W=0$. This limitation of the
Weierstrass representation may be resolved by studying the higher variations,
$\delta^3W$ and $\delta^4W,$ which is beyond the scope of the present
maniscript.

\begin{figure}[h!]\begin{center}\begin{tabular}{ccc}
\psfig{figure=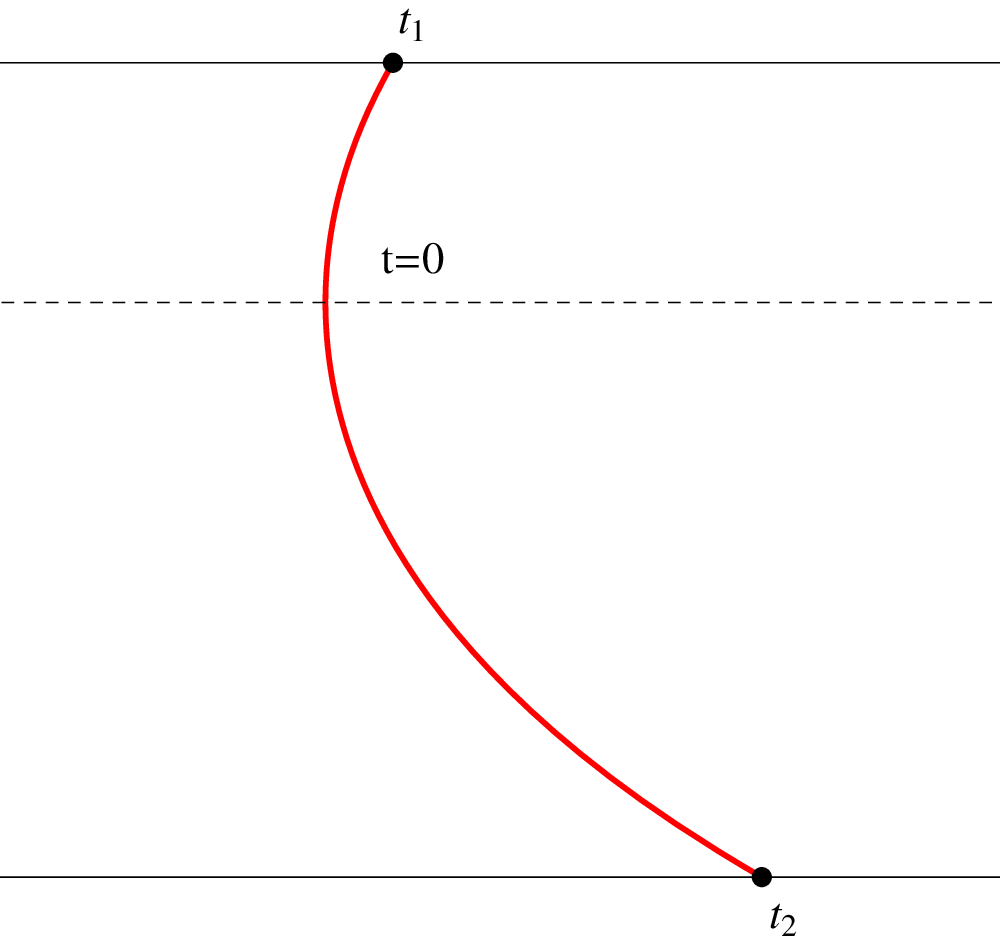,width=4cm}
\hspace{.5cm}&\hspace{.5cm}&\hspace{.5cm}
\psfig{figure=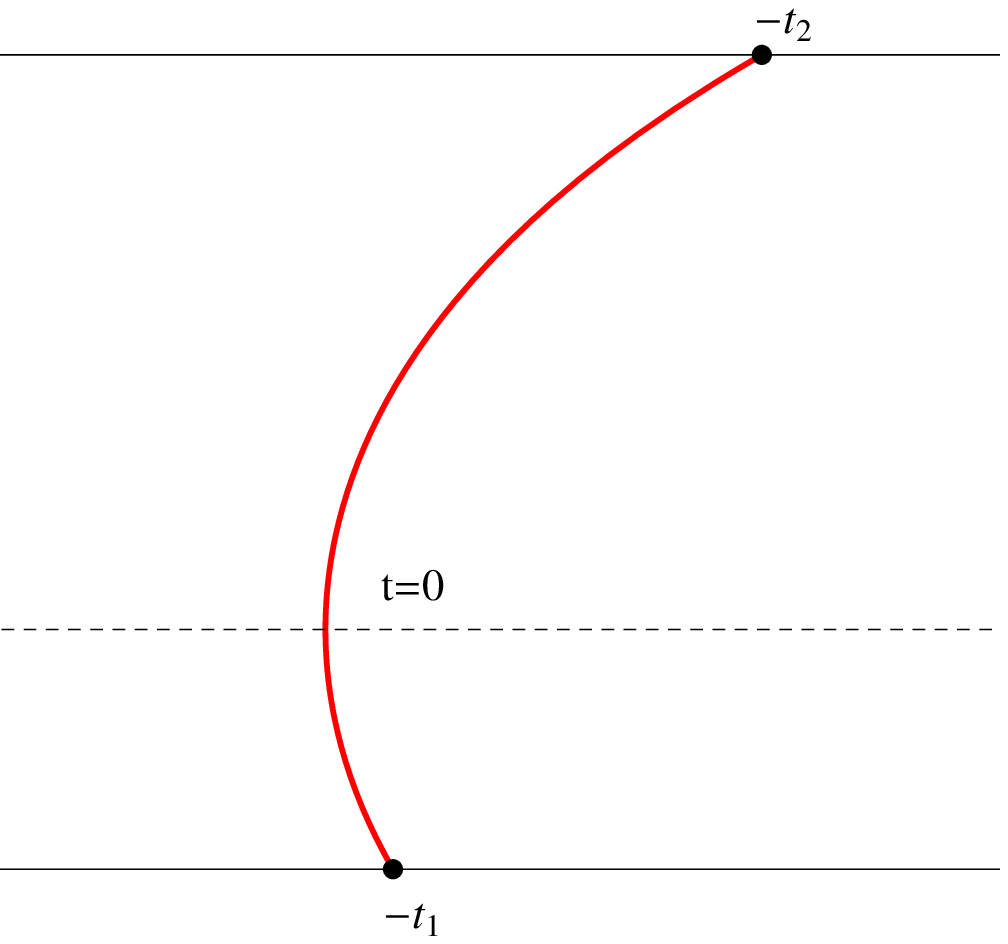,width=4cm}\\
(a) & & (b)
\end{tabular}\end{center}
\caption{Sketches of menisci between two plates showing the endpoints (a)
$t_1,t_2$ and (b) $-t_2,-t_1$.}\label{t_1_and_t_2}
\end{figure}

Consider two menisci related by symmetry reflection $t_2\to -t_1,t_1\to -t_2$
w.r.t. a midline between two solids and normal to the curve $\{\bar {x}(t),
\bar{y}(t)\}$ at the point $t=0$ (or to continuation of curve if $0\not\in
[t_2,t_1]$) as shown in Figure \ref{t_1_and_t_2}.
It is easy to conclude that the stability conditions (\ref{g16}) serve for
both menisci simultaneously,
\bea
Q_{ii}(-t_1,-t_2)=Q_{jj}(t_2,t_1),\;i\neq j=1,2;\quad Q_{12}(-t_1,-t_2)=Q_{12}
(t_2,t_1).\label{r1}
\eea
Consider a symmetric setup: $t_1=-t_2=t$, when two solid bodies are similar and
separated by a reflection plane located in the midpoint of the meniscus at $t=0$.
Then due to (\ref{r1}) the necessary conditions (\ref{g16}) 
read
\bea
Q_{11}(-t,t)=Q_{22}(-t,t)\geq 0,\quad Q_{33}(-t,t)=Q_{11}^2(-t,t)-Q_{12}^2
(-t,t)\geq 0.\label{r2}
\eea
Expression (\ref{g14}) and conditions (\ref{g16}) encompass the case when the
extremal curve is perturbed at interval $[t_2,t_1]$ including only one endpoint
(say, $t_1$) while another is left fixed. Here, instead of (\ref{g14}, \ref{g16}) we have
$\delta^2W=Q_{11}\left(\delta\tau_1\right)^2,\;Q_{11}(t_2,t_1)\geq 0.$
\section{Application to the problem of pendular rings}\label{s5}
\begin{figure}[h!]\begin{center}
\psfig{figure=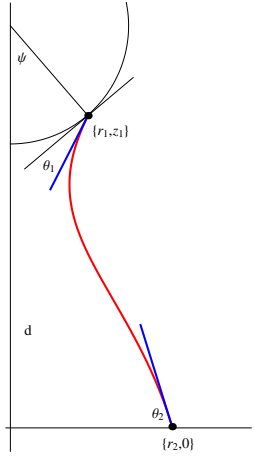,height=6cm}
\end{center}
\caption{A sketch of meniscus between plane and sphere showing the contact
angles $\theta_1,\theta_2$, filling angle $\psi$ and coordinates of the
endpoints.}\label{MeniscusSpherePlane}
\end{figure}

Apply our approach to study the stability of axisymmetric PR between solid
bodies in absence of gravity. The axial symmetry of bodies is assumed along
$z$-axis (see Figure \ref{MeniscusSpherePlane}). The shapes of meniscus $\{r(
\phi),z(\phi)\}$ and two solid bodies $\{R_j(\psi_j),Z_j(\psi_j)\}$ are given in
cylindrical coordinates, {\em i.e.}, the following correspondence holds,
\bea
x\to r,\quad y\to z,\quad X_j\to R_j,\quad Y_j\to Z_j,\quad t\to\phi,\quad\tau_j
\to\psi_j.\nonumber
\eea
The filling angle $\psi_j$ along the $j$ solid-liquid interface is chosen to
satisfy $0\leq\psi_j\leq\infty$ for unbounded solid bodies (semispace with
planar boundary, paraboloid, catenoid) and $0\leq\psi_j<\infty$ for bounded
solid bodies (sphere, prolate and oblate ellipsoids).

The functional $W$ and its integrands in (\ref{e4}) read
\bea
W\!=\!\int_{\phi_2}^{\phi_1}\!\!\!F(r,r',z,z')d\phi-\sum_{j=1}^2(-1)^j\!\!\!
\int_0^{\psi_j^*}\!\!\!G_jd\psi_j,\;\;F\!=\!\left[\gamma_{lv}\sqrt{r'^2+z'^2}-
\frac{\lambda rz'}{2}\right]r,\nonumber\\
G_j=\left[\frac{\lambda R_jZ_j'}{2}-(-1)^j(\gamma_{ls_j}-\gamma_{vs_j})
\sqrt{R_j'^2+Z_j'^2}\right]R_j,\;\label{h2}
\eea
where coefficients $\gamma_{lv}$, $\gamma_{ls_j}$ and $\gamma_{vs_j}$, $j=1,2$,
describe surface energy density at three interfaces: liquid-vapor,
solid-vapor and solid-liquid for the upper ($j=1$) and lower
($j=2$) solid bodies.
The two ELE (\ref{e14}) are reduced to a single YLE
\be
2H=\frac{z'}{r(r'^2+z'^2)^{1/2}}+\frac{z''r'-z'r''}{(r'^2+z'^2)^{3/2}},\quad
H=\frac{\lambda}{2\gamma_{lv}},\label{h3}
\ee
where $H$ stands for the meniscus mean curvature. The transversality conditions
(\ref{e15}) are known as the Young relations for the contact angle $\theta_j$
of the meniscus with the $j$-th solid body: $\cos\theta_j+(\gamma_{ls_j}-
\gamma_{vs_j})/\gamma_{lv}=0$. According to (\ref{e13}) the quantity $\eta_j=
\eta(\phi_j,\psi_j^*)$ is given by
\bea
\eta_j=\bar{z}'(\phi_j)R'(\psi_j^*)-\bar{r}'(\phi_j)Z'(\psi_j^*).\label{h4b}
\eea
Define a contact angle $\theta_j$ between meniscus and solid body as follows
\bea
\theta_j=(-1)^{j-1}\left(\arctan\frac{\bar{z}'(\phi_j)}{\bar{r}'(\phi_j)}-
\arctan\frac{Z'(\psi_j^*)}{R'(\psi_j^*)}\right),\label{h4c}
\eea
where $0\leq\arctan(\bar{z}'/\bar{r}'),\;\arctan(Z'/R')\leq\pi$. The contact 
angle $\theta_j$ vanishes when $\bar{z}'/\bar{r}'=Z'/R'$, {\em i.e.}, $\eta_j=
0$, which manifests meniscus' nonexistence at a critical angle $\phi_j^{
\bullet}$ in accordance with (\ref{f12})
\bea
\bar{z}'(\phi_j^{\bullet})R'(\psi_j^*)-\bar{r}'(\phi_j^{\bullet})Z'(\psi_j^*)=0,
\quad \bar{r}(\phi_j^{\bullet})-R(\psi_j^*)=0.\label{h4a}
\eea
Rescale the integrands in (\ref{e4}) by $2\gamma_{lv}|H|$ and deal henceforth
with expressions,
\bea
F\!=\!\left[\sqrt{r'^2+z'^2}-\frac{S_H}{2}rz'\right]\!r,\;G_j\!=\!\left[\frac{
S_H}{2}R_jZ_j'+(-1)^j\cos\theta_j\sqrt{R_j'^2+Z_j'^2}\right]\!R_j,\;\label{h5}
\eea
where $S_H\!=\!\mbox{sign}H$. Straightforward calculation in (\ref{e21}) gives
an expression for $K_j$,
\bea
K_j=U_j\eta_j,\quad U_j=-\frac{R_j}{2\sqrt{\bar{r}_j'^2+\bar{z}_j'^2}}
\left(\frac{\bar{z}_j''R'_j-\bar{r}_j''Z_j'}{\bar{r}_j'^2+\bar{z}_j'^2}-
\frac{Z_j''R'_j-R_j''Z'_j}{R_j'^2+Z_j'^2}\right),\label{h5a}
\eea
where $\bar{f}_j'\!=\!\bar{f}'(\phi_j)$, $\bar{f}_j''\!=\!\bar{f}''(\phi_j)$.
Combining (\ref{h5a}, \ref{g15}) write expressions for $Q_{ij}(\phi_2,\phi_1)$,
\bea
Q_{11}=\eta_1\left(\frac{\eta_1P_{11}}{2\Delta}+U_1\right),\quad
Q_{22}=\eta_2\left(\frac{\eta_2P_{22}}{2\Delta}-U_2\right),\quad
Q_{12}=\frac{\eta_1\eta_2P_{12}}{2\Delta},\label{h5b}
\eea
that results in $Q_{33}\propto\eta_1\eta_2$ and according to (\ref{r1}) we have
$U_j(-\phi,\psi^*)=U_j(\phi,\psi^*)$. Thus, stability domain ${\sf Stab}_2(
\phi_1,\phi_2)$ of liquid meniscus of any type has boundaries including meniscus
nonexistence lines $\phi_j=\phi_j^{\bullet}$ given by (\ref{h4a}).

Find formulas for $H_j$ in (\ref{f4}) by substituting (\ref{h5}) into
(\ref{e18}, \ref{f2}) and obtain
\bea
H_1\!=\!\frac{\bar{r}}{\left(\bar{r}'^2+\bar{z}'^2\right)^{3/2}},\;\;H_2\!=\!
\frac{\left(H_1\bar{r}''\right)'}{\bar{r}'},\;\;H_3\!=\!\bar{r},\;(H_1w')'\bar{
r}'-\left(H_1\bar{r}''\right)'w\!=\!\mu\bar{r}'\bar{r}.\;\;\label{h6}
\eea
Fundamental solutions of equation (\ref{h6}) read,
\bea
\bar{w}_1=\bar{r}'(\phi),\quad\bar{w}_2=E(\phi)\bar{r}'(\phi),\quad E(\phi)=
g\int^{\phi}\!\!\!\!\frac{dt}{H_1\bar{r}'^2}=g\!\!\int^{\phi}\!\!\frac{\left(
\bar{r}'^2+\bar{z}'^2\right)^{3/2}dt}{\bar{r}'^2\bar{r}}.\nonumber
\eea
\subsection{Pendular rings with zero curvature}\label{s51}
For $H=0$ the first Delaunay's type, {\em catenoid} (${\sf Cat}$) appears from 
(\ref{h3}),
\bea
\bar{r}=\sec\phi,\quad \bar{z}=\ln\frac{\cos\phi}{1-\sin\phi}+C,\quad\frac{
\bar{z}'}{\bar{r}'}=\cot\phi,\quad\bar{r}'^2+\bar{z}'^2=\bar{r}^4,\label{h8}
\eea
where $C$ is the constant determined from the BC. Entries in (\ref{e17}) read,
\bea
H_1\!=\!\frac{1}{\bar{r}^5},\;\;H_2\!=\!-4\frac{\bar{r}^2-1}{\bar{r}^5},\;\;
L\!=\!\frac{\bar{r}^3\bar{r}'-\bar{z}'\bar{z}''}{\bar{r}^5},\;\;N\!=\!-\frac{
\bar{r}'\bar{r}''}{\bar{r}^5},\;\;M\!=\!\frac{\bar{z}'\bar{r}''}{\bar{r}^5}.
\label{h9}
\eea
Note that $H_1(\phi)$ is always positive, therefore the set ${\mathbb L}(\phi_1,
\phi_2)$ is given by the whole lower halfplane $\{\phi_2<\phi_1\}$. The Jacobi
equation (\ref{h6}) in this case reads
\bea
w''-5w'\tan\phi+4w\tan^2\phi=\mu\sec^6\phi.\nonumber
\eea
Its fundamental and particular solutions and auxiliary functions read,
\bea
&&\bar{w}_1=\tan\phi\;\sec\phi,\quad\bar{w}_2=\sec^2\phi-T(\phi)\bar{w}_1,\quad
T(\phi)=\ln\left(\tan\phi+\sec\phi\right),\nonumber\\
&&\bar{w}_3=-\frac{\sec^4\phi}{2}+\frac{3}{4}\bar{w}_1\left[T(\phi)+\bar{w}_1
\right],\quad I_2(\phi)=\frac{T(\phi)}{4}\left[3-4I_1(\phi)\right]+\frac{3}{4}
\bar{w}_1,\nonumber\\
&&I_1(\phi)=\frac{\sec^2\phi}{2},\quad I_3(\phi)=\frac{3T(\phi)}{32}\left[
8I_1(\phi)-5\right]+\frac{\bar{w}_1}{32}\left[4I_1(\phi)-15\right].\label{h12}
\eea
The determinant $\Delta_{cat}(\phi_1,\phi_2)=\Delta_{cat}$ is given by
\bea
\frac{32\Delta_{cat}}{K_{12}}\!=\!T_{12}(7M_3\!-\!2M_5\!-\!6M_1)\!-\![3T_{12}
^2\!-\!3L_2\!+\!4L_4]J_{12}\!+\!(L_2\!-\!2)(2L_2\!-\!3)K_{12},\nonumber
\eea
where $T_{12}=T(\phi_1)-T(\phi_2)$, $\;J_{12}=
\tan\phi_1\tan\phi_2$, $\;K_{12}
=\sec\phi_1\sec\phi_2,$ and
\bea
L_n=\sec^n\phi_1+\sec^n\phi_2,\quad M_n=\tan
\phi_1\sec^n\phi_2-\tan\phi_2\sec^n\phi_1.\nonumber
\eea
Matrix elements $P_{ij}$ calculated from (\ref{g15})
are too cumbersome to be presented here. Functions $\eta(\phi_j,\psi_j^*)$ and
$K(\phi_j,\psi_j)$ are calculated substituting (\ref{h8}) into
(\ref{h4b}, \ref{h5a}).
\subsubsection{${\sf Cat}$ meniscus between two plates}\label{s511}
\begin{figure}[h!]\begin{center}\begin{tabular}{ccc}
\psfig{figure=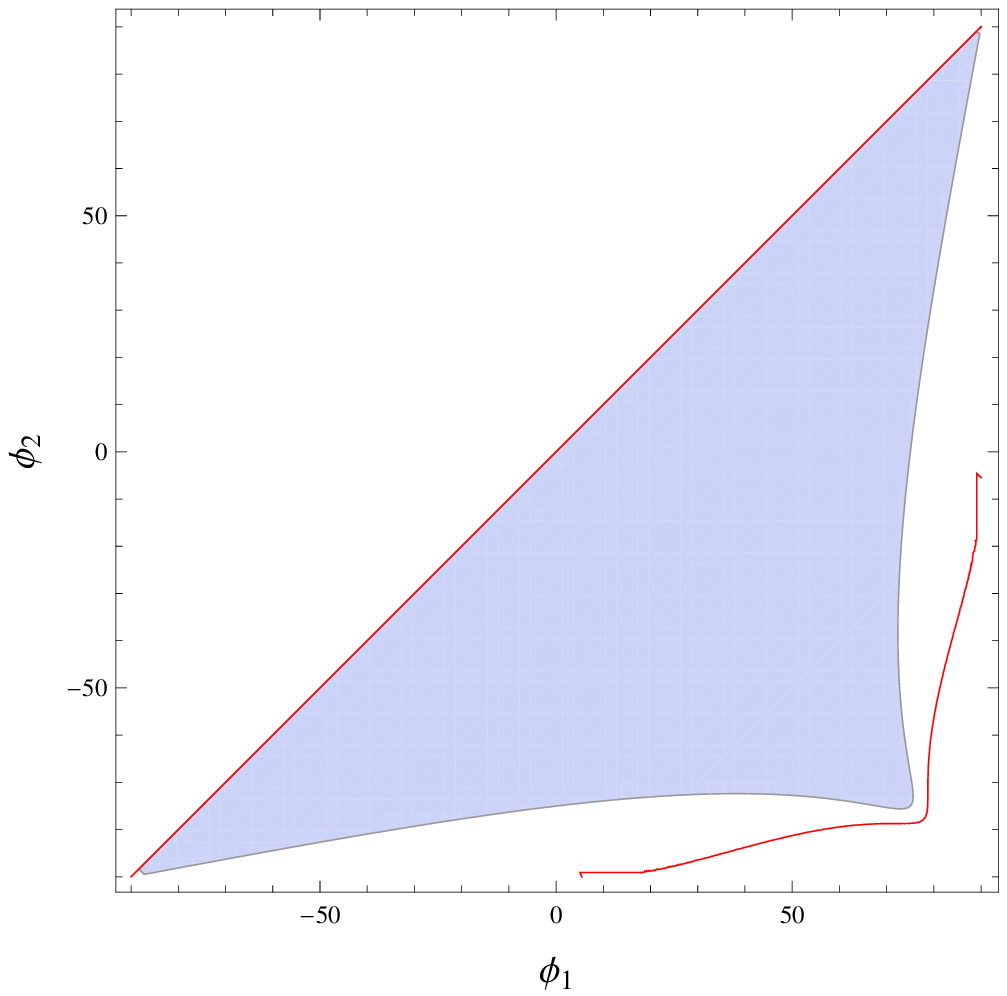,width=5cm}
\hspace{.01cm}&\hspace{.01cm}&\hspace{.01cm}
\psfig{figure=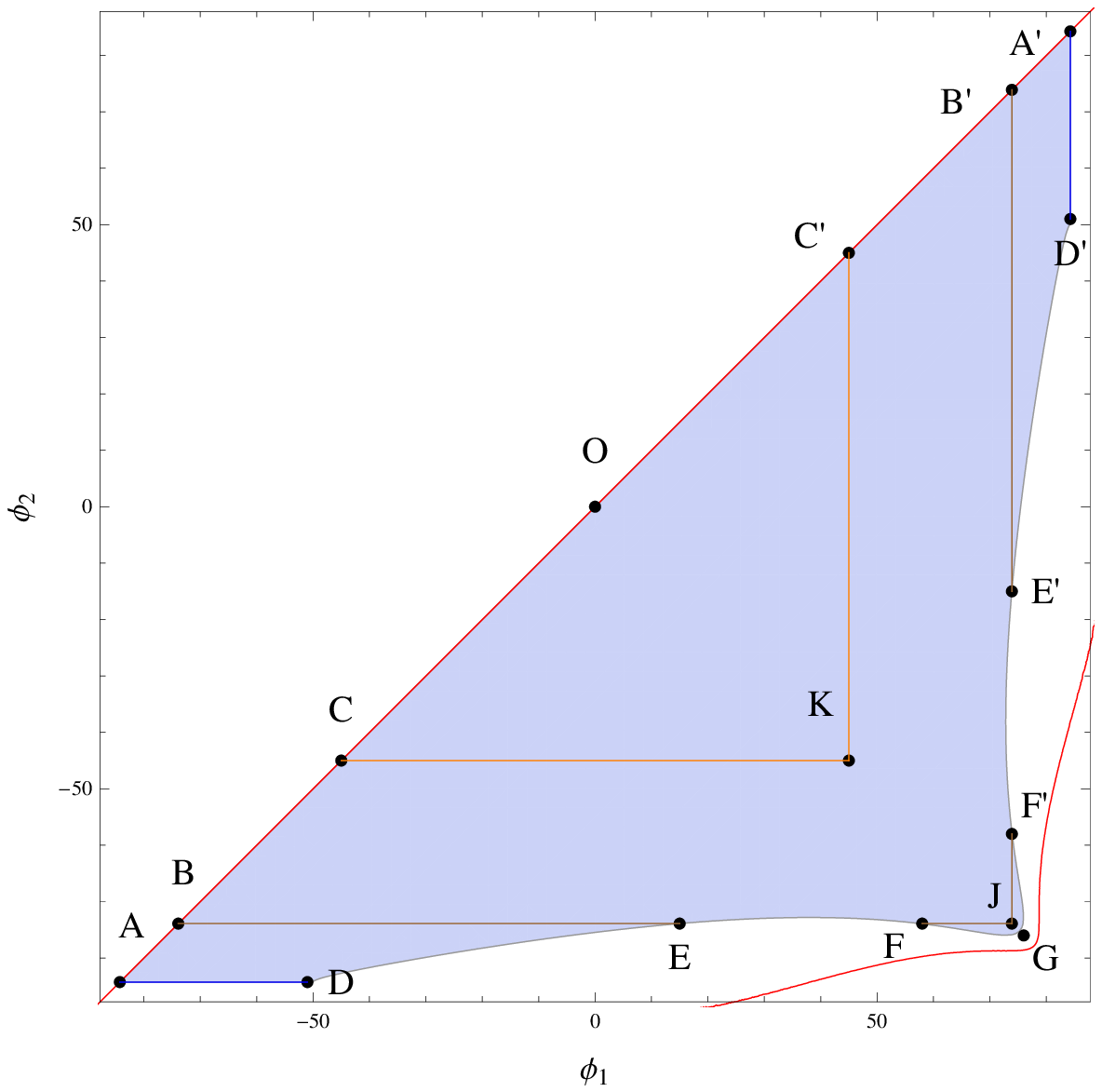,
width=5cm}\\
(a) & & (b)
\end{tabular}\end{center}
\caption{(a) Stability diagram ({\em SD}) for ${\sf Cat}$ menisci between two
plates in a halfplane $\phi_1>\phi_2$ is shaded in {\em gray}. (b) The {\em SD}
for ${\sf Cat}$ menisci between two equal spheres are represented by interiors
of polygons: $A=100$, \{OCBADEFGF'E'D'A'B'C'O\}, $\phi^{\bullet}(1,100)=84.3^o$;
$A=13$, \{OCBEFJF'E'A'B'C'O\}, $\phi^{\bullet}(1,13)=73.9^o$; $A=4$, \{OCKC'O\}
$\phi^{\bullet}(1,4)=60^o$. The {\em red curves} show the location of conjugate
points while the {\em blue lines} show the location of points where $\eta(\phi_j
^{\bullet},\psi_j^*)=0$, $\sec\phi_j^{\bullet}=100\sin\psi_j^*$.}
\label{Cat_2_plane}
\end{figure}

The ${\sf Cat}$ with given endpoints on two solid plates exists for arbitrary
contact angles $\theta_j$. Parametrization of plates and relations between
$\phi_j$ and $\theta_j$ read (see Figure \ref{Cat_2_plane}(a))
\bea
R_j\!=\!A\psi_j,\;\;Z_j\!=\!d_j,\;\;\theta_j\!=\!\frac{\pi}{2}+(-1)^j\phi_j,\;\;
\eta_j\!=\!A\sec\phi_j,\;\;2K_j\!=\!-A^2\sin\phi_j\cos^2\phi_j.\nonumber
\eea
By (\ref{h4a}) the critical angles $\phi_j^{\bullet}$ read: $\phi_j^{\bullet}=
(-1)^{j+1}\pi/2$, that makes every point of infinite
plates (at the distance $d=d_1-d_2$) attainable by ${\sf Cat}$ meniscus.

In Figure \ref{Cat_2_plane}(a) the red curve determines the boundaries of
${\sf Stab}_1(\phi_1,\phi_2)$ defined in (\ref{f13}) while the lower boundary
of stability domain gives the boundaries of ${\sf Stab}_2(\phi_1,\phi_2)$
defined in (\ref{g18}). Numerical calculations show a nice coincidence with
boundaries found in the framework of Vogel's approach in \cite{Zho97},
\bea
5\int_{\phi_1}^{\phi_2}\cos^{-5}t\;dt\cdot\int_{\phi_1}^{\phi_2}\cos^{-1}t\;dt=
9\left(\int_{\phi_1}^{\phi_2}\cos^{-3}t\;dt\right)^2.\label{h16}
\eea
In symmetric setup (\ref{r2}) ${\sf Cat}$ meniscus between two plates is stable
if $\theta\geq 14.97^o$.
\vspace{-.2cm}
\subsubsection{${\sf Cat}$ meniscus between two ellipsoids}\label{s512}
Consider ${\sf Cat}$ meniscus between two axisymmetric ellipsoids given by
equation $R_j^2+\left(Z_j-g_j\right)^2\epsilon_j^{-2}=A^2$, $\epsilon_j>0$,
where $\epsilon_j$ stands for anisotropy parameter and $\{0,g_j\}$ denotes
coordinates of the $j$-th ellipsoid center. Ellipsoids may be specified as
prolate ($\epsilon_j>1$) and oblate ($\epsilon_j<1$). The upper and lower
ellipsoids are separated by distance $d=g_1-g_2-A(\epsilon_1+\epsilon_2)$ and
given parametrically,
\bea
R_j\!=\!A\sin\psi_j,\;\;Z_j\!=\!g_j\!+\!(-1)^{j}A\epsilon_j\cos\psi_j,\quad\eta
_j\!=\!\frac{\sqrt{A^2\cos^2\phi_j-1}+(-)^j\epsilon_j\tan\phi_j}{\cos^2\phi_j},
\nonumber\\
K_j=-\frac{\eta_j\cos\phi_j}{2}\left(\eta_j\sin\phi_j\cos^3\phi_j+(-1)^j
\epsilon_j\left[1+\frac1{\cos^2\phi_j+\epsilon_j^2\sin^2\phi_j}\right]\right).
\hspace{1cm}\nonumber
\eea
According to (\ref{h4c}) the contact angles are given by
$$\theta_j=\frac{\pi}{2}+(-1)^j\phi_j-\arctan\frac{\epsilon_j}
{\sqrt{A^2\cos^2\phi_j-1}}.
$$
By (\ref{h4a}) the critical angles $\phi_j^{\bullet}=\phi_j^{\bullet}
(\epsilon_j,A)$ are given by equation,
\be
A^2\cos^4\phi_j^{\bullet}+(\epsilon_j^2\!-\!1)\cos^2\phi_j^{\bullet}\!-\!
\epsilon_j^2\!=\!0,\quad\phi_1^{\bullet}(1,A)\!=\!-\phi_2^{\bullet}(1,A)\!=\!
\arccos\frac1{\sqrt{A}},\quad\label{h18}
\ee
where $\epsilon_j=1$ stands for two equal spheres. This makes the areas,
attainable by ${\sf Cat}$ stable meniscus on the spheres, substantially
limited. Figure \ref{Cat_2_plane}(b) shows stability diagrams ({\em SD}) of 
${\sf Cat}$ menisci between two equal spheres of different radii. Decrease 
of $A$ reduces the stability domain ${\sf Stab}_2(\phi_1,\phi_2)$ caused 
by non-planar solid bodies and decrease of $\phi_j^{\bullet}$. For $A<11.7$ 
the domain ${\sf Stab}_2(\phi_1,\phi_2)$ is a right isosceles triangle 
$\{OCKC'O\}$, otherwise the domain has curvilinear boundaries.
\subsubsection{${\sf Cat}$ meniscus between other solid bodies}\label{s513}
The theory of PR stability with free CL developed in section \ref{s4} can be
applied to arbitrary pair of axisymmetric solid bodies. Here we study another
pair, two paraboloids. Consider the ${\sf Cat}$ meniscus between two convex
parts of axisymmetric solid bodies,
\be
R_j=A\psi_j,\quad Z_j=g_j+(-1)^{j+1}AC_ja_j(\psi_j/a_j)^{\nu_j},\quad a_j,
\;\nu_j,\;C_j,\;A>0.\label{h19}
\ee
\begin{figure}[h!]\begin{center}\begin{tabular}{ccc}
\psfig{figure=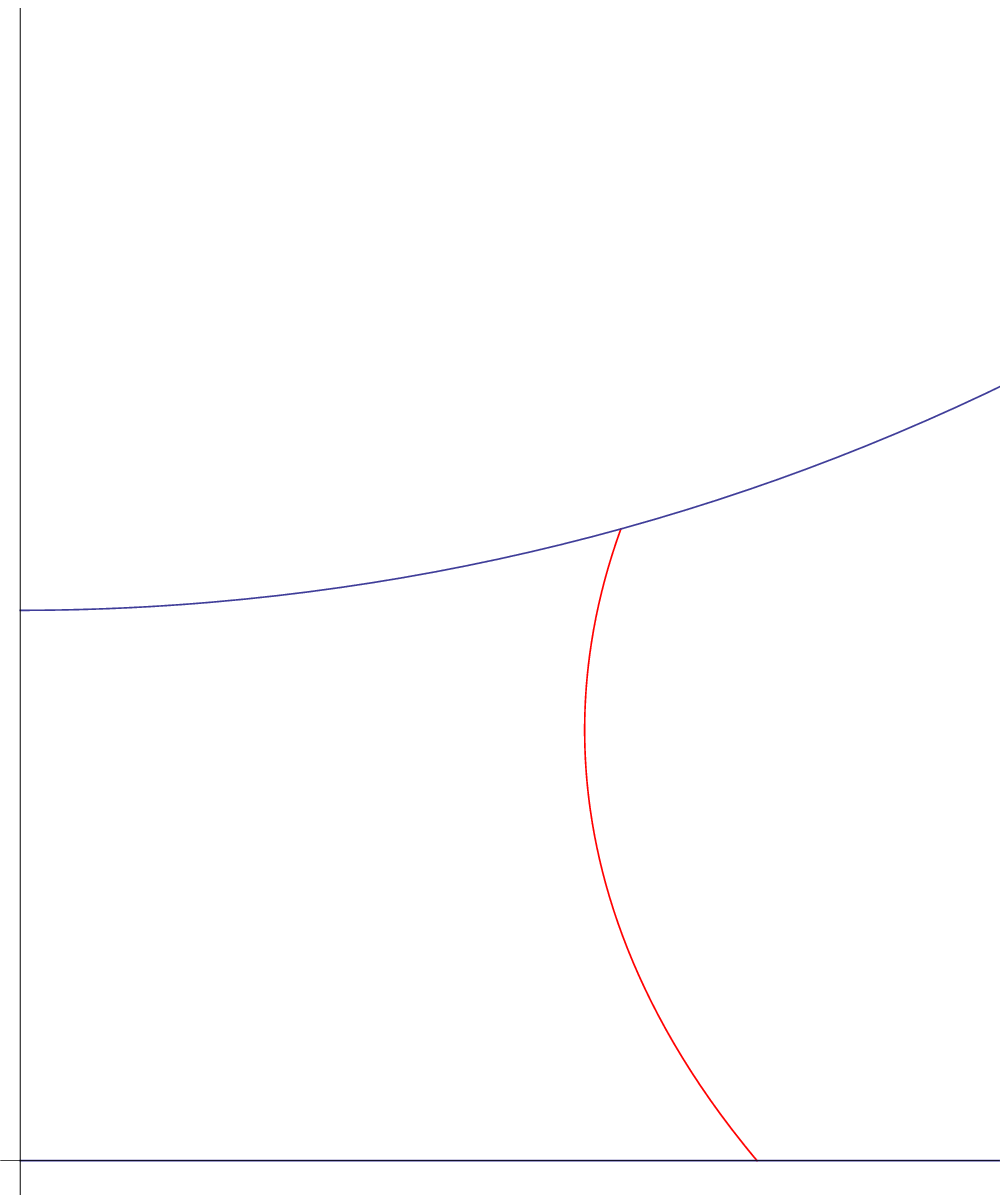,height=
3cm}\hspace{.01cm}&\hspace{.01cm}
\psfig{figure=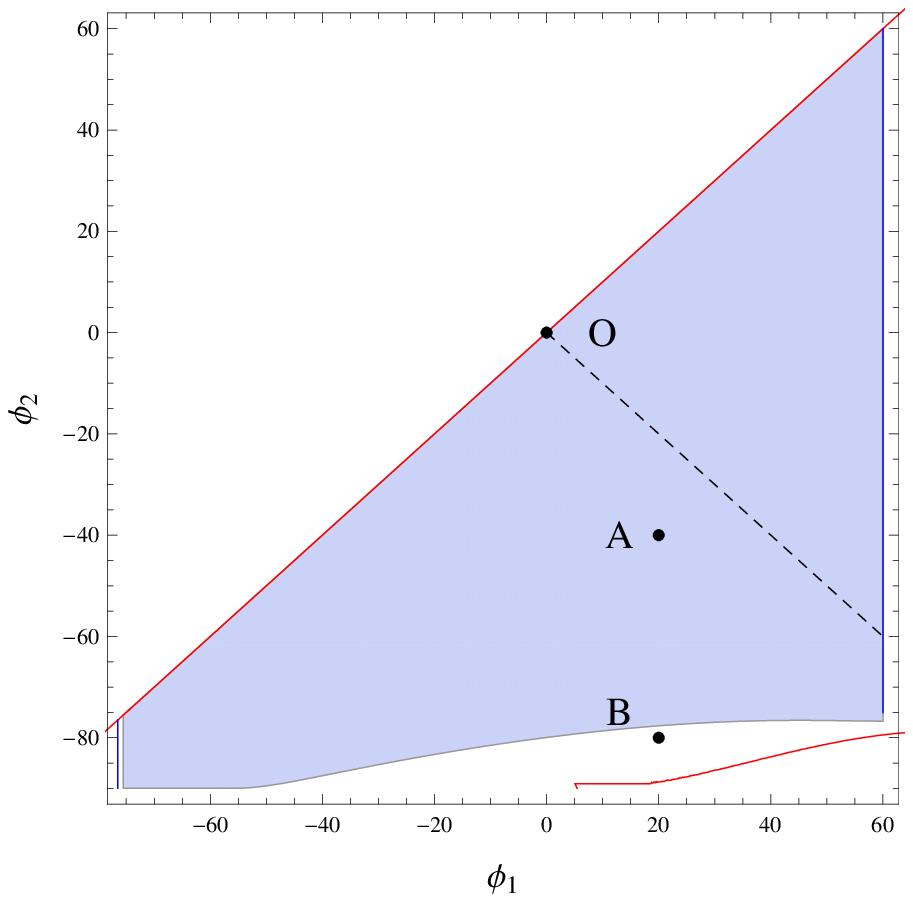,
height=4.5cm}\hspace{.01cm}&\hspace{.01cm}
\psfig{figure=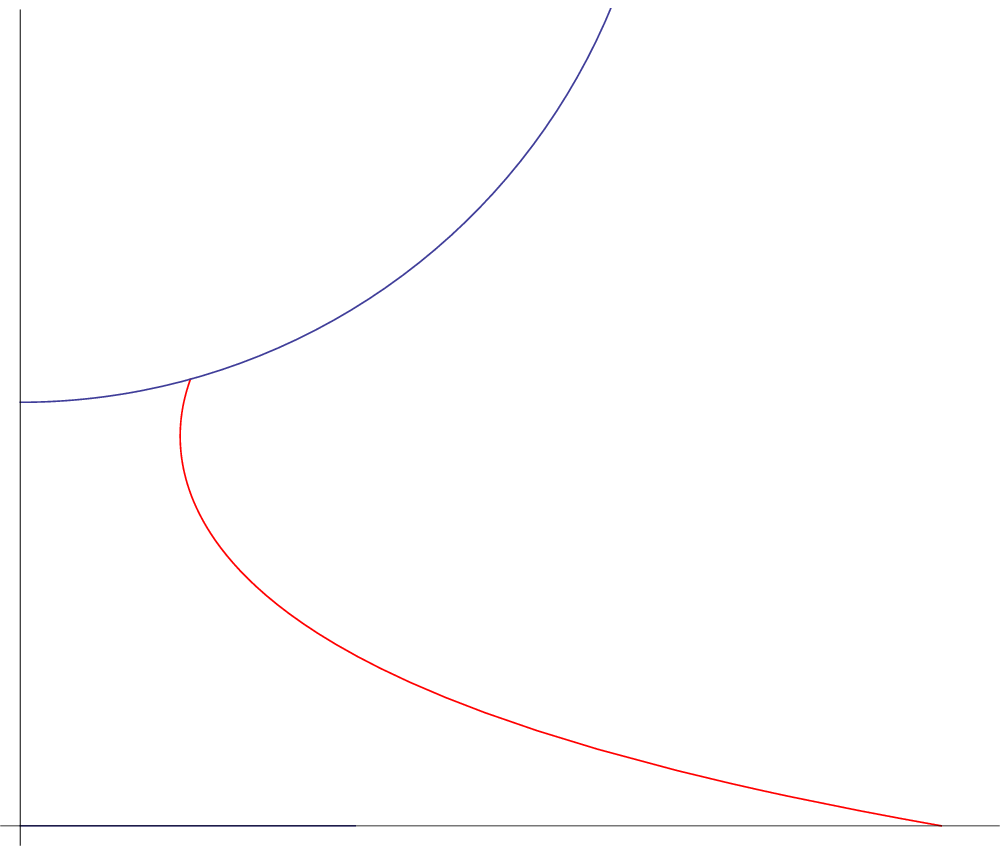,height=
3cm}\\
(a) & (b) & (c) \\
\end{tabular}\end{center}
\caption{The {\em SD} (b) for ${\sf Cat}$ menisci between solid plate and
sphere, $A=4$, is not symmetric w.r.t. the {\em dashed} line $\phi_1+\phi_2=0$.
The critical angles are $\phi_1^{\bullet}(1,4)=60^o$, $\phi_2^{\bullet}(0,4)=
-75.5^o$. Points $A$ and $B$ mark (a) stable $\phi_1=20^o,\phi_2=-40^o,$ and
(c) unstable $\phi_1=20^o,\phi_2=-80^o,$ menisci, respectively.}
\label{sph_plan_cat}
\end{figure}
For $\nu_j>1$ the surface is smooth at $\psi_j=0$, otherwise it has a
singularity point. The case $\nu_j=1$ represents a conic surface. The critical
angles $\phi_j^{\bullet}$ are given by relations,
\bea
\nu_jC_j\tan\phi_j^{\bullet}=\left(a_jA\cos\phi_j^{\bullet}\right)^{\nu_j-1},
\;\nu_j>1;\quad\cot\phi_j^{\bullet}=C_j,\;\nu_j=1.\nonumber
\eea
When ${\sf Cat}$ meniscus connects solid bodies of different shape the stability
domain loses its symmetry w.r.t. the line $\phi_1+\phi_2=0$, thus breaking an
equality $\phi_1^{\bullet}=-\phi_2^{\bullet}$ for critical angles. This can be
seen in the setup of meniscus between solid sphere and plate at Figure
\ref{sph_plan_cat}, for which according to (\ref{h18}) we have,
$$
\phi_1^{\bullet}(1,A)=\arccos\frac1{\sqrt{A}},\quad
\phi_2^{\bullet}(0,A)=-\arccos\frac1{A}.
$$
\section{Pendular rings with nonzero curvature} \label{s6}
For $H\neq 0$ the equation (\ref{h3}) is solved in elliptic integrals of the 
first $F$ and the second $E$ kind. Here we choose a parametrization similar to 
that used in \cite{Mlad2007},
\bea
\bar{r}(\phi)=\sqrt{1+B^2+2B\cos\phi},\quad\bar{z}(\phi)=M(\phi,B)-M(\phi_2,B)+
Z_2(\psi_2),\quad\label{k1}\\
M(\phi,B)=(1+B)E\left(\phi/2,m\right)+(1-B)F\left(\phi/2,m\right),\quad
m^2=\frac{4B}{(1+B)^2},\nonumber
\eea
where $m$ stands for modulus of elliptic integral. The expression for $B$ is 
given by
$$
B^2+2B\cos\phi_1+1=R_1^2(\psi_1).
$$
The solution derivatives satisfy the relationships
\bea
\frac{\bar{r}'}{B}=-\frac{\sin\phi}{\bar{r}},\;\;\frac{\bar{r}''}{B}=\frac{\bar{
r}'\sin\phi}{\bar{r}^2}-\frac{\cos\phi}{\bar{r}},\;\;\bar{z}'=\frac{1+B\cos\phi}
{\bar{r}},\;\;\bar{z}''=\frac{\bar{r}'(\bar{r}-\bar{z}')}{\bar{r}}.\;\;
\label{k2}
\eea
Formulas (\ref{k1}) describe four Delaunay's types \cite{Delaunay1841} of
surfaces of revolution with constant $H$: {\em cylinder} (${\sf Cyl}$), $B=0$,
{\em unduloid} (${\sf Und}$), $B<1$, {\em sphere} (${\sf Sph}$), $B=1$, and
{\em nodoid} (${\sf Nod}$), $B>1$. Entries in (\ref{e17}) read,
\bea
H_1=H_3=\bar{r},\;\;H_2=-\left(\bar{r}+2\bar{r}''\right),\;\;L=\bar{r}'-\bar{z}'
\bar{z}''\bar{r},\;\;N=-\bar{r}'\bar{r}''\bar{r},\;\;M=\bar{z}'\bar{r}''\bar{r}.
\nonumber
\eea
Note that $\bar{r}'^2+\bar{z}'^2=1,$ and $H_1$ is positive as in section
\ref{s51}. Equation (\ref{h6}) reads
\be
w''-\frac{B\sin\phi}{\bar{r}^2}w'+\left(1-\frac{2B\cos\phi}{\bar{r}^2}-
\frac{2B^2\sin^2\phi}{\bar{r}^4}\right)w=\mu.\label{k5}
\ee
Its fundamental and particular solutions and corresponding auxiliary functions
read:
\bea
&&\bar{w}_1=\frac{\sin\phi}{\bar{r}},\quad\bar{w}_2=\cos\phi+(1+B)M_1\bar{w}_1,
\quad \bar{w}_3=1+(1+B)M_2\bar{w}_1,\nonumber\\
&&I_1=-\cos\phi,\quad\eta_j=\frac1{\bar{r}(\phi_j)}\left[(1+B\cos\phi_j)R_j'
(\psi_j^*)+B\sin\phi_jZ_j'(\psi_j^*)\right],\nonumber\\
&&I_2\!=\!\bar{r}\sin\phi+(1+B)(I_1M_1+M_2),\;I_3\!=\!(1+B)\left[2E\left(\frac{
\phi}{2},m\right)+I_1M_2+M_1\right]\quad\nonumber\\
&&M_1(\phi,m)=E\left(\frac{\phi}{2},m\right)-F\left(\frac{\phi}{2},m\right)+
M_2,\quad M_2(\phi,m)=\frac{m^2}{2}F\left(\frac{\phi}{2},m\right).
\nonumber
\eea
Expression for $\Delta(\phi_1,\phi_2)$ for arbitrary meniscus of nonzero
curvature is too long to be presented here.
\subsection{Stability of cylinder menisci ${\sf Cyl}$}\label{s61}
Specify the above formulas for ${\sf Cyl}$ meniscus,
\bea
&&B=0,\;\bar{r}=1,\;\bar{z}=\phi,\;\bar{w}_1=\sin\phi,\;\bar{w}_2=\cos\phi,\;
\bar{w}_3=1,\;L=M=N=0,\quad\nonumber\\
&&I_1=-\cos\phi,\quad I_2=\sin\phi,\quad I_3=\phi,\quad\eta_j=R_j'(\psi_j^*),
\quad K_j=\xi_j.\label{k8}
\eea
Expressions for $\Delta_{Cyl}(\phi_1,\phi_2)$ and matrix elements $P_{ij}$ read
\bea
&&\Delta_{Cyl}(\phi_1,\phi_2)=\Delta\phi\;\Gamma_1\left(\frac{\Delta\phi}{2}
\right)\sin\Delta\phi,\quad\Gamma_1(x)=1-\frac{\tan x}{x},\quad\Delta\phi=
\phi_1-\phi_2,\nonumber\\
&&P_{11}=P_{22}=\Delta\phi\;\Gamma_1\left(\Delta\phi\right)\cos\Delta\phi,\quad
P_{12}=-\Delta\phi\Gamma_2(\Delta\phi),\quad\Gamma_2(x)=1-\frac{\sin x}{x}.
\nonumber
\eea
\subsubsection{${\sf Cyl}$ meniscus between two plates}\label{s611}
We have $\theta_1=\theta_2=\pi/2$ and $R_j=\psi_j$, $Z_j=d$, $K_j=0,$ leading to
\bea
Q_{11}\!=\!Q_{22}\!=\!\frac{\Gamma_1\left(\Delta\phi\right)}{\Gamma_1\left(
\Delta\phi/2\right)}\cot\Delta\phi,\;\;Q_{12}\!=\!-\frac{\Gamma_2\left(\Delta
\phi\right)}{\Gamma_1\left(\Delta\phi/2\right)}\csc\Delta\phi,\;\;Q_{33}\!=\!-
\frac1{\Gamma_1\left(\Delta\phi/2\right)}.\quad\nonumber
\eea
There are no conjugate points in region $\Delta_{Cyl}(\phi_1,\phi_2)<0$, {\em 
i.e.}, $\Delta\phi<2\pi$. The stability domains ${\sf Stab}(\Delta\phi)$ for 
three different BCs are the following
\bea
&&(a)\;\mbox{fixed endpoints}:\;\;\Delta_{Cyl}<0\;\Rightarrow\;0<\Delta\phi<2
\pi,\nonumber\\
&&(b)\;\mbox{one endpoint is free and another is fixed}:\;Q_{11}>0\;\Rightarrow
0<\Delta\phi<\varkappa\pi,\nonumber\\
&&(c)\;\mbox{free endpoints}:\;\;Q_{33}>0\;\Rightarrow\;0<\Delta\phi<\pi,
\nonumber
\eea
where $\varkappa=\min\{x_*\;|\;\tan x_*=x_*,\;x_*>0\}\simeq 1.4303$. Stability
of ${\sf Cyl}$ meniscus between two plates is well studied and often compared 
\cite{Vog1987}, \cite{Langbein02} to the Plateau-Rayleigh instability of a slow 
flowing liquid jet of infinite length. Its threshold coincides with the case 
(a) above in the following  sense: the jet of the circular cross-section is 
stable if the length of fluctuations does not exceed the circumference.
\subsubsection{${\sf Cyl}$ meniscus between two ellipsoids or plate and
ellipsoid}\label{s612}
Using parametrization of section \ref{s512} allow anisotropy $\epsilon$ to get
both positive and negative values that distinguishes the exterior (convex) 
ellipsoid shape ($\epsilon>0$) and its interior (concave, or hollow) shape 
($\epsilon<0$),
\bea
\frac{Q_{jj}}{A^2}=\frac{\epsilon_j\sin\psi_j^*\cos\psi_j^*}{\epsilon_j^2\sin^2
\psi_j^*+\cos^2\psi_j^*}+\frac{P_{jj}\cos^2\psi_j^*}{\Delta_{Cyl}},\quad
\frac{Q_{12}}{A^2}=\frac{P_{12}\cos\psi_1^*\cos\psi_2^*}{\Delta_{Cyl}},\nonumber
\eea
where $P_{ij}$ are given in section \ref{s61}. Consider a case of ${\sf Cyl}$
between equal ellipsoids.
\begin{figure}[ht!]\begin{center}\begin{tabular}{cc}
\psfig{figure=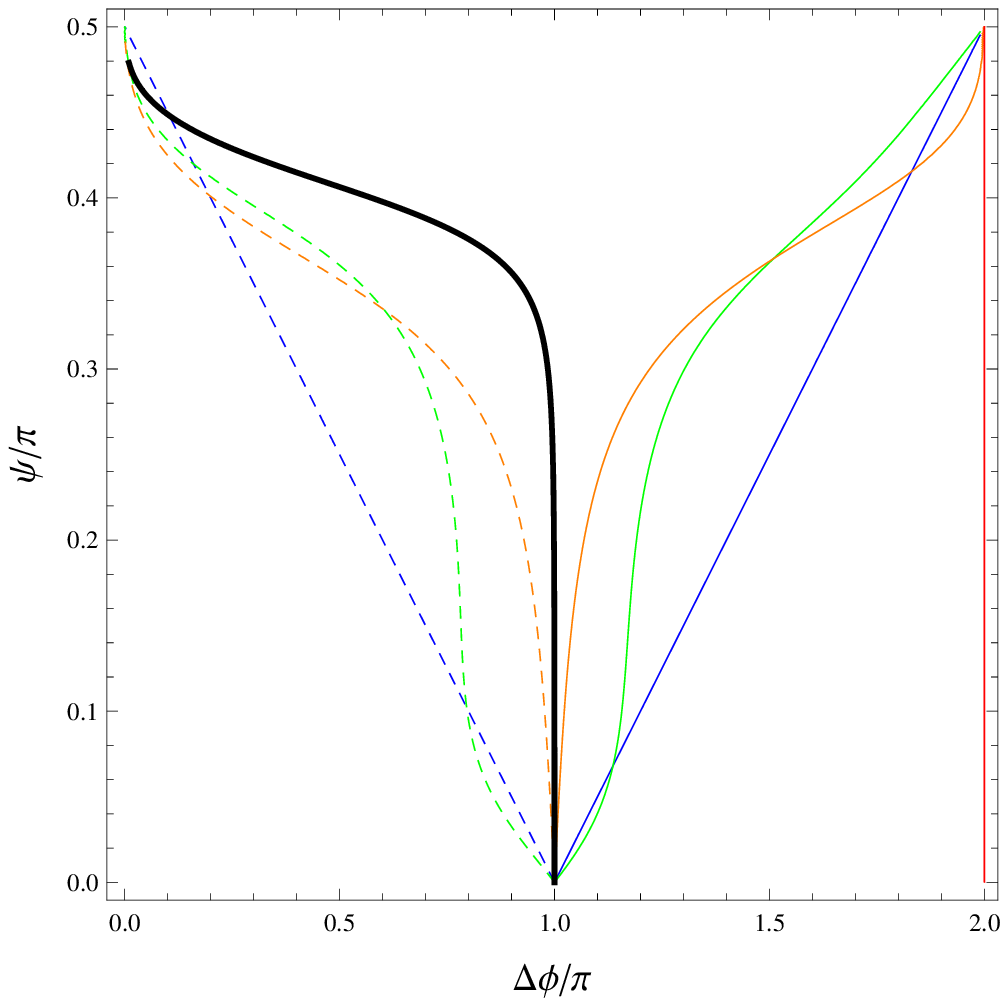,width=4.5cm}&
\psfig{figure=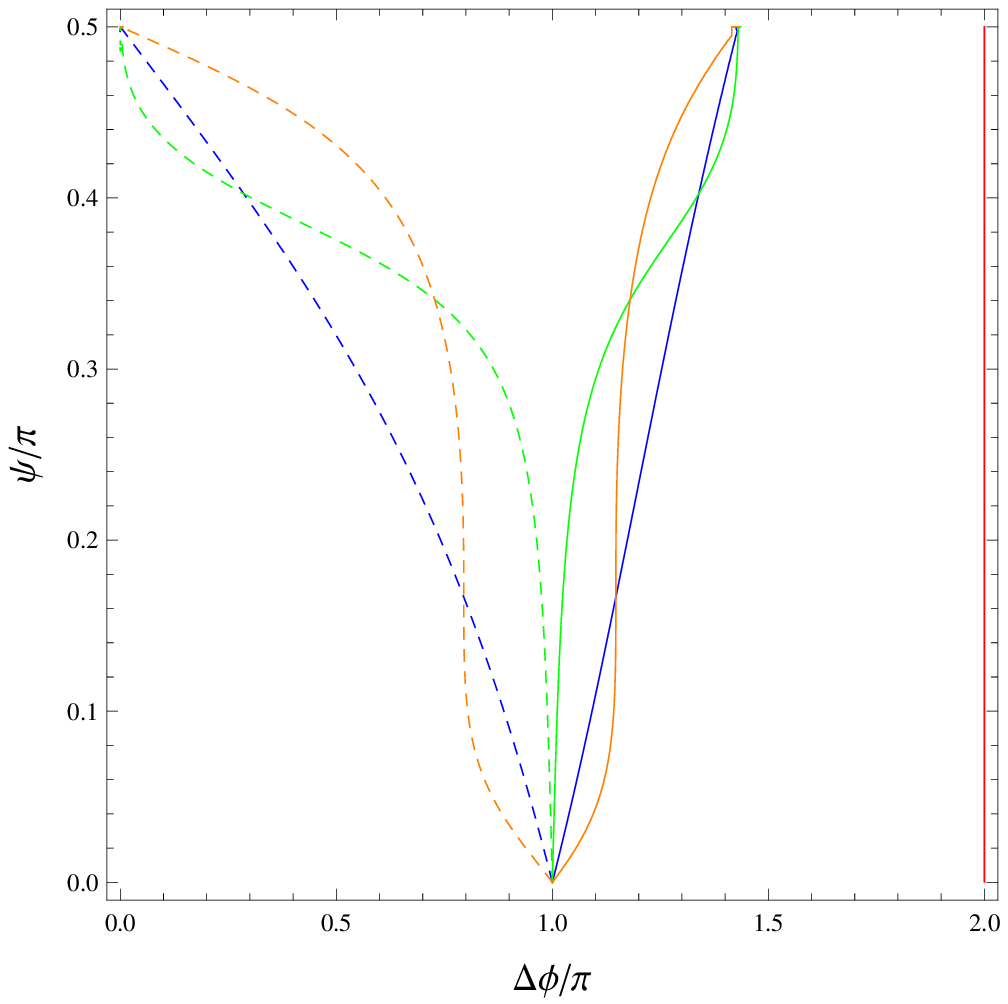,width=4.5cm}\\
(a) & (b) \\
\end{tabular}\end{center}
\caption{(a) The right boundaries of {\em SD} for ${\sf Cyl}$ menisci between
two solid ({\em plain}) and hollow ({\em dashed}) ellipsoids shown in {\em
blue}: $\epsilon_j\!=\!1(-1)$, {\em green}: $\epsilon_1\!=\!3(-3),\;\epsilon_2
\!=\!0.1(-0.1),$ and {\em orange}: $\epsilon_1\!=\!0.05(-0.05),\;\epsilon_2\!=
\!0.15(-0.15)$. The {\em thick black curve} corresponds to ${\sf Cyl}$ meniscus
between solid and hollow ellipsoids ($\epsilon_1\!=\!-\epsilon_2=\!0.05$). (b) 
The right boundaries of {\em SD} for ${\sf Cyl}$ menisci between plate and 
convex ({\em plain}) or hollow ({\em dashed}) ellipsoids shown in {\em blue}: 
$\epsilon_1\!=\!\epsilon_2\!=\!1(-1),$ {\em orange}: $\epsilon_1\!=\!\epsilon_2
\!=\!3(-3),$ and {\em green}: $\epsilon_1\!=\!\epsilon_2\!=0.1(-0.1)$. The 
left boundary of {\em SD} in both Figures (a,b) coincides with the $\psi$ axis.}
\label{diag_ell}
\end{figure}
The stability criteria (\ref{g15}) give rise to the {\em SD} boundaries by 
equation,
\bea
\cot\frac{\Delta\phi}{2}+\frac{\epsilon\tan\psi^*}{\epsilon^2\sin^2\psi^*+
\cos^2\psi^*}=0,\label{k12}
\eea
that results in solutions for spheres (see Figure \ref{diag_ell}(a), for
$\epsilon=1$ it coincides with that of reported in \cite{Vog1999},
\bea
\psi^*=\frac{-\pi+\Delta\phi}{2},\;\;1\leq\frac{\Delta\phi}{\pi}\leq 2\quad
\mbox{and}\quad\psi^*=\frac{\pi-\Delta\phi}{2},\;\;0\leq\frac{\Delta\phi}
{\pi}\leq 1.\nonumber
\eea
The case of ${\sf Cyl}$ meniscus between the plate and ellipsoid gives,
\bea
\frac{Q_{11}}{A^2}=\frac{\epsilon\sin\psi^*\cos\psi^*}{\epsilon^2\sin^2\psi^*+
\cos^2\psi^*}+\frac{P_{11}\cos^2\psi^*}{\Delta_{Cyl}},\quad\frac{Q_{22}}{A^2}=
\frac{P_{22}}{\Delta_{Cyl}},\quad\frac{Q_{12}}{A^2}=\frac{P_{12}\cos\psi^*}
{\Delta_{Cyl}}.\nonumber
\eea
Its stability is governed by equation,
\bea
\frac{\tan(\Delta\phi)}{\Gamma_1(\Delta\phi)}-\frac{\epsilon\tan\psi^*}
{\epsilon^2\sin^2\psi^*+\cos^2\psi^*}=0,\label{k14}
\eea
that results in solutions for sphere ($\epsilon=1$) upon the plate (see Figure
\ref{diag_ell}(b)),
\bea
\cot\psi^*=\cot\Delta\phi-\frac1{\Delta\phi},\;1\leq\frac{\Delta\phi}{\pi}
\leq\varkappa,\quad\cot\psi^*=\frac1{\Delta\phi}-\cot\Delta\phi,\;0\leq\frac{
\Delta\phi}{\pi}\leq 1.\nonumber
\eea
\subsubsection{${\sf Cyl}$ meniscus between two paraboloids or two catenoids}
\label{s613}
\begin{figure}[h!]\begin{center}\begin{tabular}{cc}
\psfig{figure=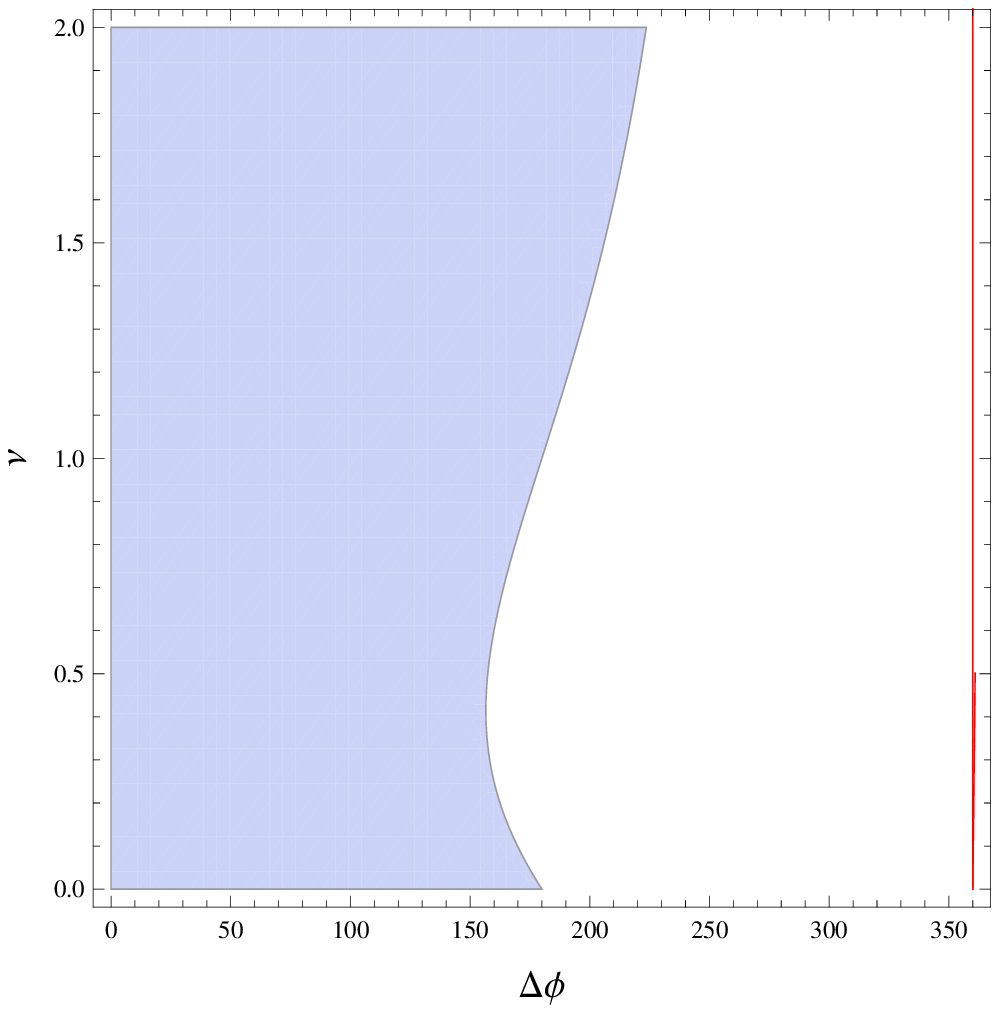,width=4.5cm}&
\psfig{figure=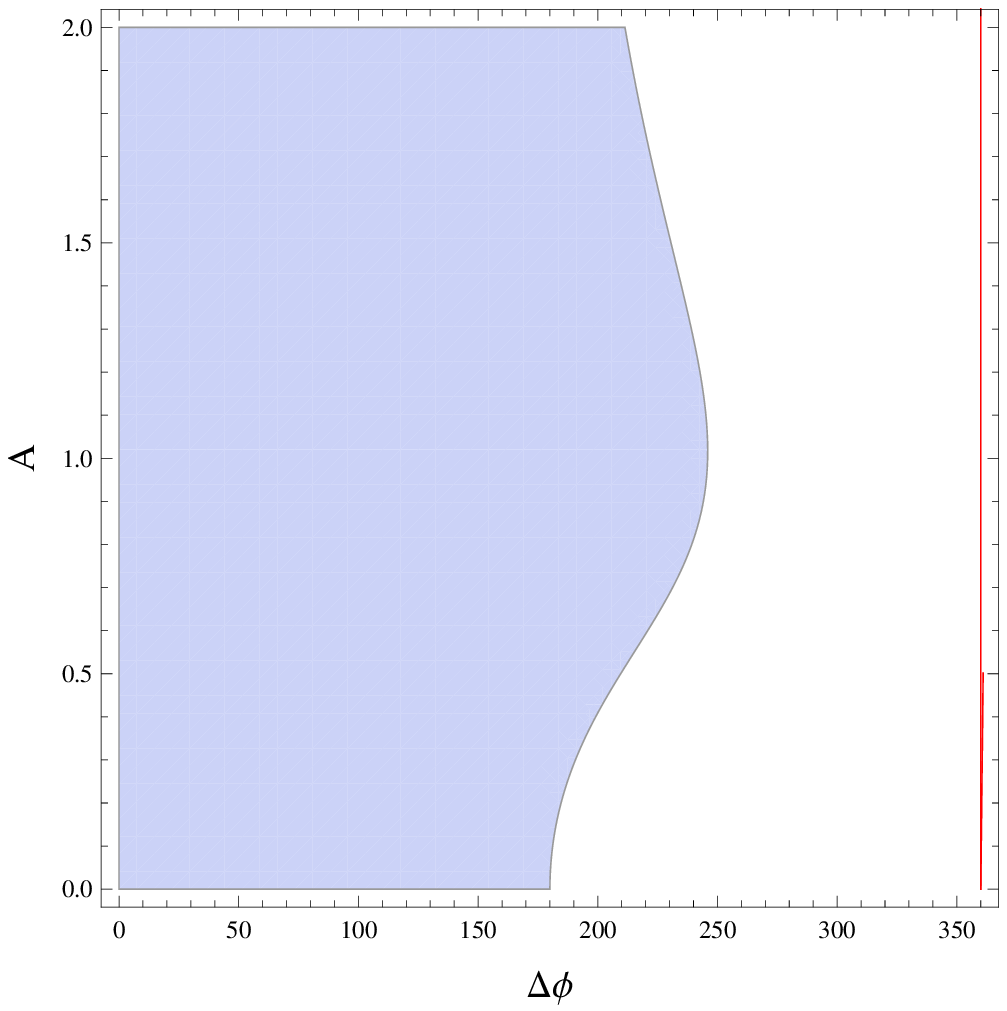,width=4.5cm}\\
(a) & (b)
\end{tabular}\end{center}
\caption{The {\em SD} for ${\sf Cyl}$ meniscus between (a) two solid paraboloids
for $a_i=C_i=1,\;\nu_1=\nu_2=\nu,$ and (b) two solid catenoids for $C_i=1,\;
b_1=b_2=b$.}\label{diag_par_cat}
\end{figure}

Using parametrization (\ref{h19}) write a matrix $Q_{ij}$ and the governing
equation for stability of ${\sf Cyl}$ between two equal paraboloids, $C_i\!=\!
C$, $a_i\!=\!a$, $\nu_i\!=\!\nu$ (see Figure \ref{diag_par_cat}(a)),
\bea
\frac{Q_{jj}}{A^2}=\rho\frac{\nu-1}{1+\rho^2}+\frac{P_{jj}}{\Delta_{Cyl}},\quad
\frac{Q_{12}}{A^2}=\frac{P_{12}}{\Delta_{Cyl}},\quad\rho=\frac{C\nu}{a^{\nu-1}},
\quad\cot\frac{\Delta\phi}{2}+\rho\frac{\nu-1}{1+\rho^2}=0.\nonumber
\eea
The ${\sf Cyl}$ meniscus between two solid catenoids,
\bea
R_j=A\psi_j,\quad Z_j=g_j+(-1)^{j+1}AC_j\cosh (b_j\psi_j),\quad C_j,b_j,A>0,
\label{k17}
\eea
in the case of equal catenoids, $C_j=C$, $b_j=b$, produces (see Figure
\ref{diag_par_cat}(b))
\bea
\frac{Q_{jj}}{A^2}\!=\!\frac{Cb^2\cosh b}{1+C^2b^2\sinh^2b}+\frac{P_{jj}}
{\Delta_{Cyl}},\;\frac{Q_{12}}{A^2}\!=\!\frac{P_{12}}{\Delta_{Cyl}},\;\cot
\frac{\Delta\phi}{2}+\frac{Cb^2\cosh b}{1+C^2b^2\sinh^2b}\!=\!0.\nonumber
\eea
\subsection{Stability of nonzero curvature menisci between two plates}
\label{s62}
In a variety of axisymmetric menisci with $H\neq 0$ between two solid bodies
we focus on the simple case of two plates and present ${\sf Stab}_2(\phi_1,
\phi_2)$ for all
\begin{figure}[h!]\begin{center}\begin{tabular}{cc}
\psfig{figure=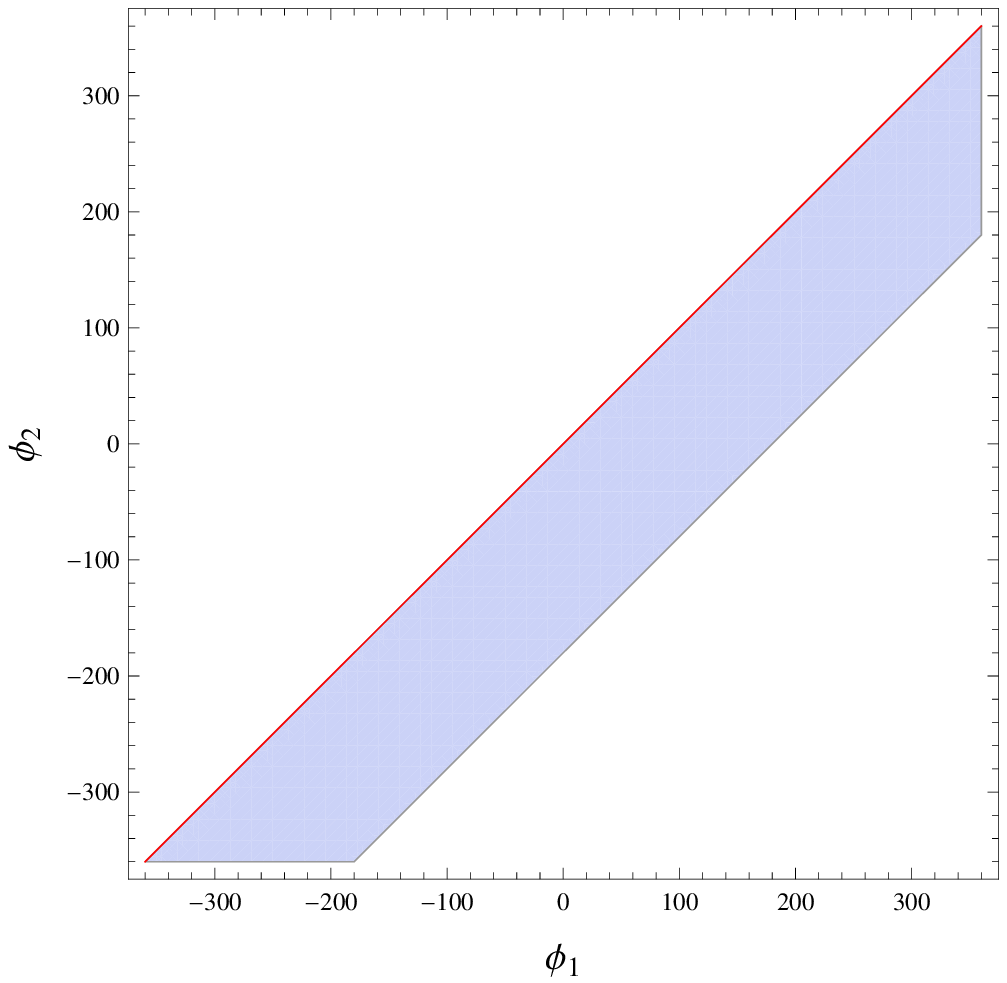,width=4.5cm}&
\psfig{figure=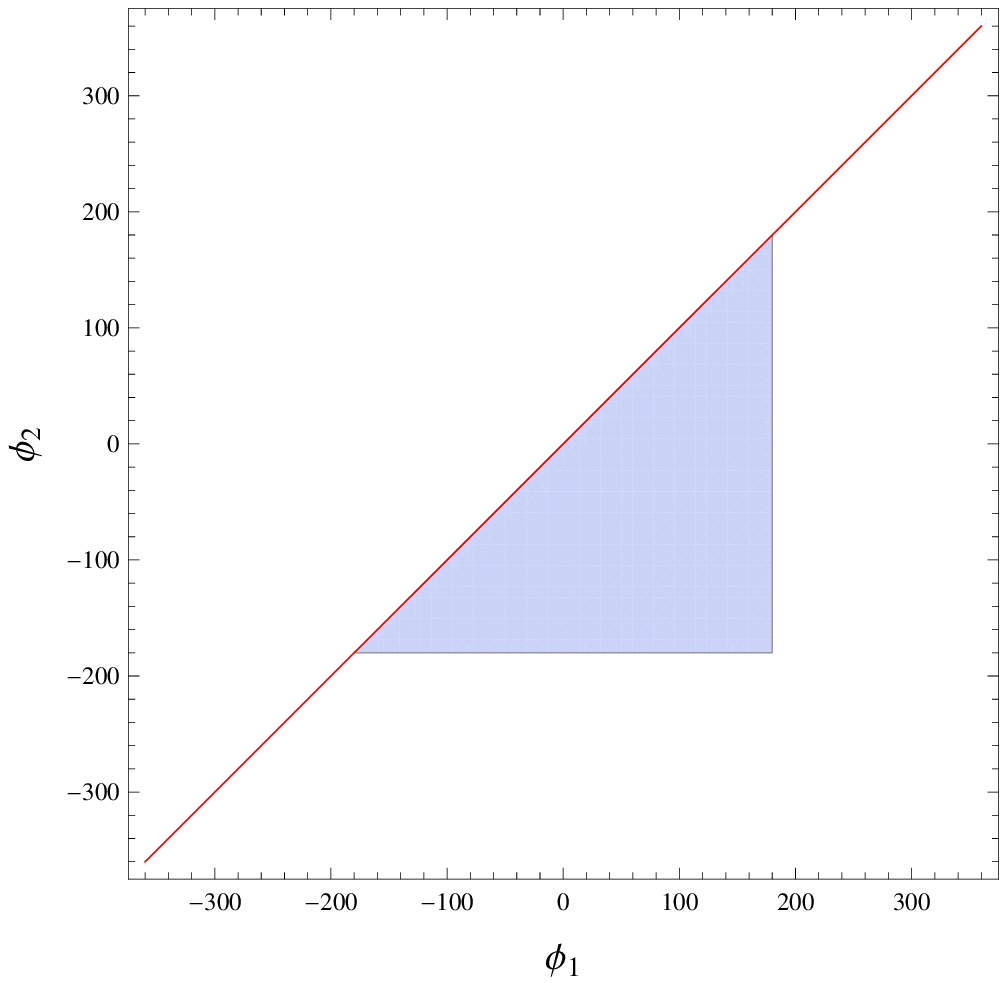,width=4.5cm}\\
(a) & (b)\\
\psfig{figure=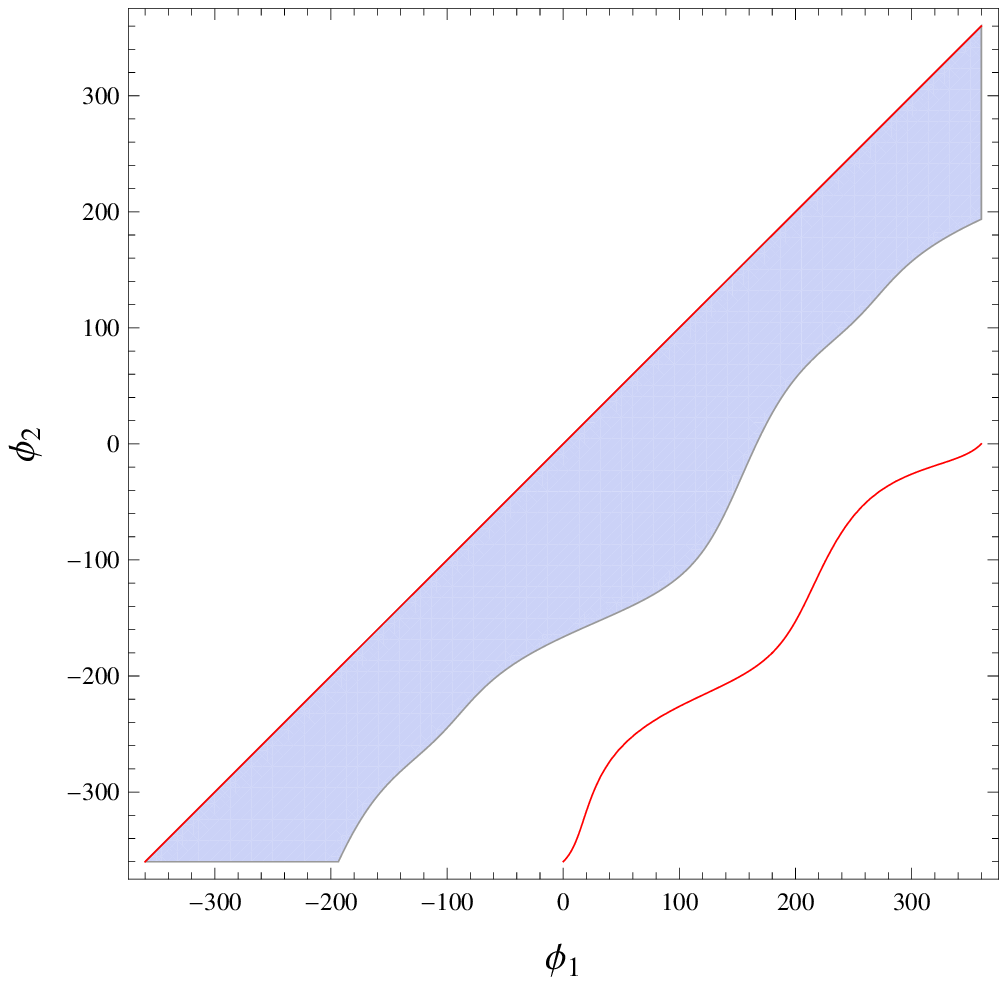,width=4.5cm}&
\psfig{figure=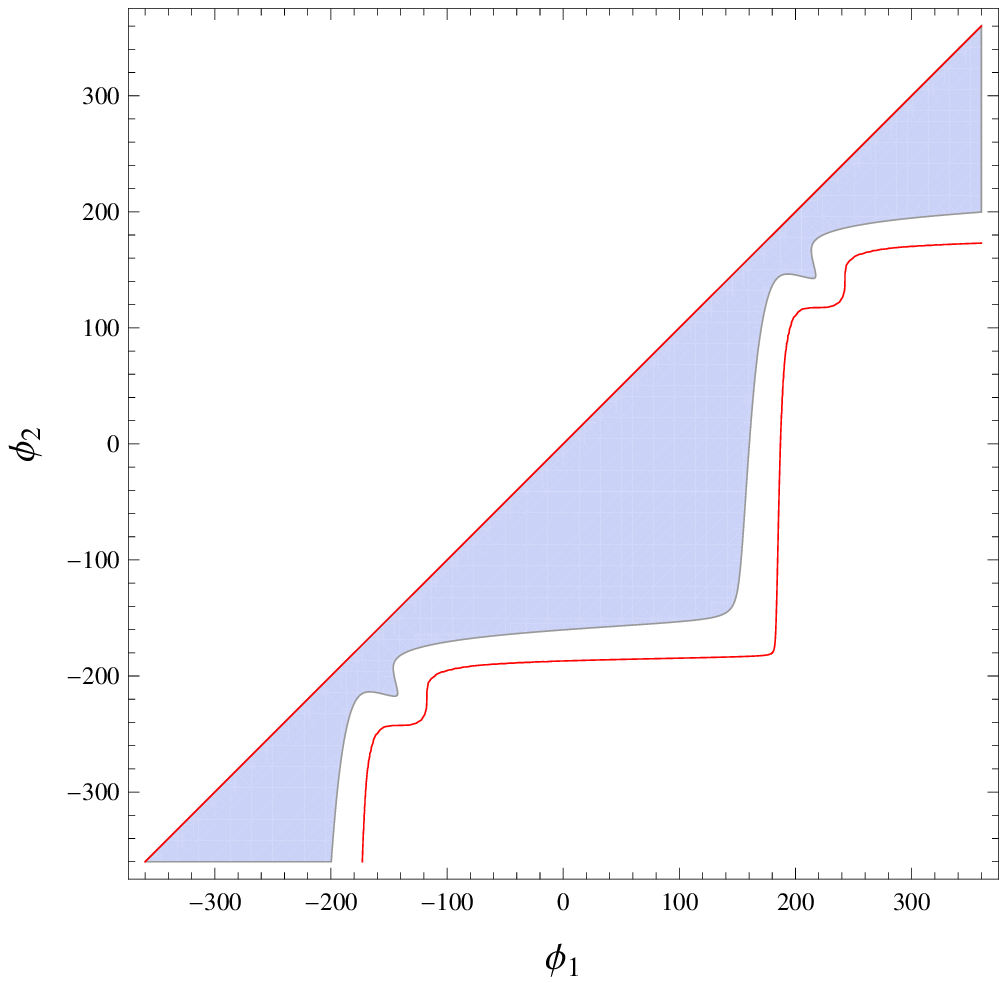,width=4.5cm}\\
(c) & (d)\\
\end{tabular}\end{center}
\caption{The {\em SD} for (a) ${\sf Cyl}$, $B=0$, (b) ${\sf Sph}$, $B=1$, and
two ${\sf Und}$ menisci, (c) $B=0.3$ and (d) $B=0.8$, between two plates. The
red curves in (c.d) show the location of conjugate points.}\label{C_S_U}
\end{figure}
menisci types. An importance of the two plates setup is based on the statement 
\cite{FinVog1992}: {\em every stable connected configuration is rotationally 
symmetric, i.e.}, axisymmetric PR between two plates under 3D non-axisymmetric 
perturbations do not bifurcate to any stable 3D non-axisymmetric PR. The 
stability triangle for ${\sf Sph}$ menisci in Figure \ref{C_S_U} (b) describes 
a single ${\sf Sph}$ segment trapped between two plates. Its right corner $\phi
_1=-\phi_2=180^o$ corresponds to the whole sphere with contact angles $\theta_1
=\theta_2=\pi$ embedded between two plates. The {\em SD} for ${\sf Und}$ menisci
in Figure \ref{C_S_U} (c,d) are intermediate domains in the range $0<B<1$ 
between ${\sf Cyl}$ and ${\sf Sph}$ menisci. The existence of {\em IP} in the 
${\sf Und}$ meridional profile ${\mathcal M}_U$ is governed by requirement:
\bea
\phi_2\leq\phi_U^{ip}\leq\phi_1,\;\;\bar{z}'(\phi_U^{ip})\bar{r}''(\phi_U^{ip})
-\bar{z}''(\phi_U^{ip})\bar{r}'(\phi_U^{ip})=0\;\Rightarrow\;\cos\phi_U^{ip}=
-B.\nonumber
\eea
A value $\phi_U^{ip}$ has important property, namely, from (\ref{g16}) we obtain
\bea
Q_{33}(\phi_U^{ip},-\phi_U^{ip})=0.\label{k20}
\eea
In section \ref{s621} we give detailed discussion of $\phi_U^{ip}$ relationship
to ${\sf Und}$ stability.

\begin{figure}[h!]\begin{center}\begin{tabular}{cc}
\psfig{figure=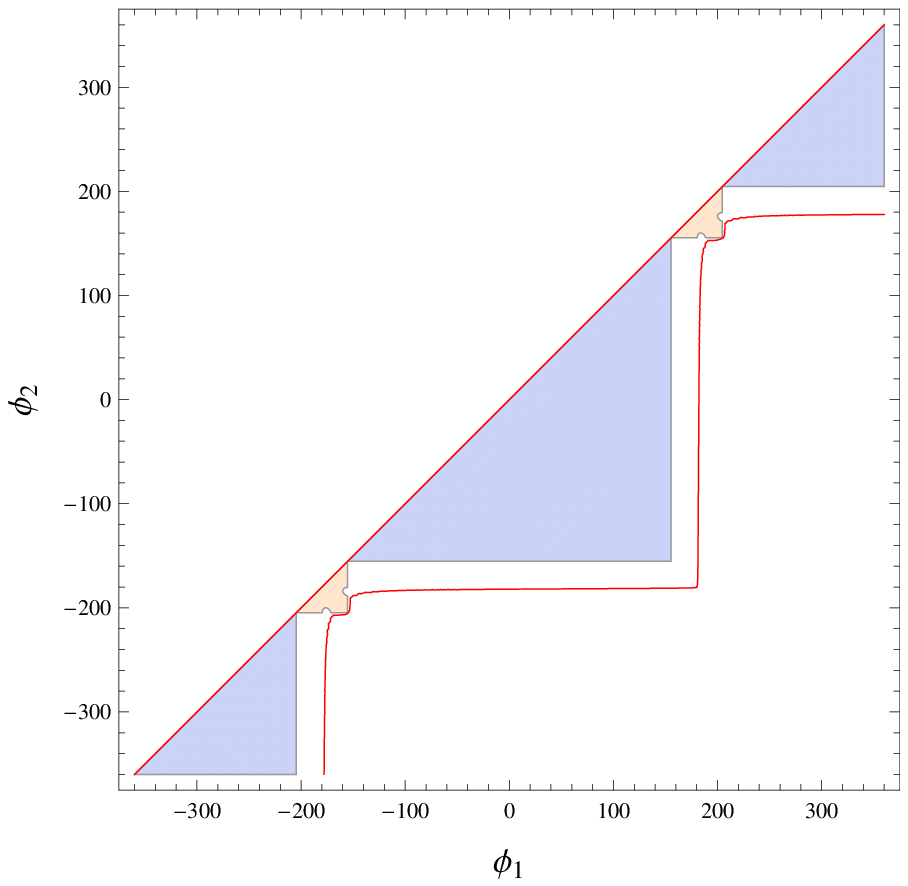,width=4.5cm}&
\psfig{figure=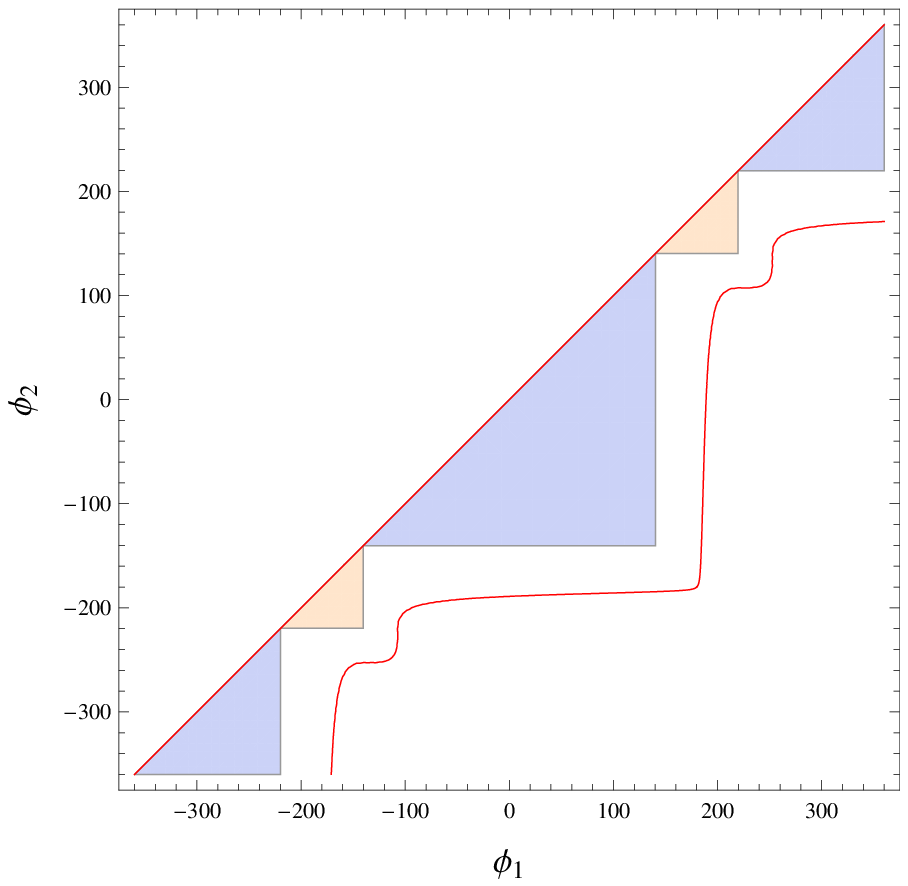,width=4.5cm}\\
(a) & (b)
\end{tabular}\end{center}
\caption{The {\em SD} for ${\sf Nod}$ menisci between two plates, with (a) $B=
1.1$ and (b) $B=1.3$. Different types of ${\sf Nod}$ menisci curvature are 
shown in {\em violet-blue} (positive) and {\em orange} (negative) colors.}
\label{Nd_and_Nd}
\end{figure}

The {\em SD} for ${\sf Nod}$ menisci in Figure \ref{Nd_and_Nd} differs from the
rest of diagrams and comprise two different sort of sub-diagrams: ${\sf Nod}$
menisci with convex and concave meridional profiles ${\mathcal M}_N$. The
positive curvature $H$ corresponds to the convex part of ${\mathcal M}_N$, while
the negative $H$ produces its concave segment. This justifies the non-existence
of ${\sf Nod}$ meniscus with both its convex and concave parts which meet at
$\phi_N^{ip}$ such that $z'(\phi_N^{ip})=0$, {\em i.e.},
$\cos\phi_N^{ip}=-B^{-1}$.
\subsubsection{${\sf Und}$ menisci with inflection point between two plates}
\label{s621}
In this section we verify three statements \cite{Ath1987}, \cite{Vog1987},
\cite{Vog1989}, \cite{FinVog1992} about stability of ${\sf Und}$ menisci with
free contact points between two plates with contact angles $\theta_1,\theta_2$.
We also present a new statement summarizing our investigations on stability
domain.

{\bf 1}. {\em If $\;\theta_1=\theta_2=\pi/2$ the ${\sf Und}$ menisci are 
unstable \cite{Ath1987}, \cite{Vog1987}}.

\begin{figure}[h!]\begin{center}\begin{tabular}{ccc}
\multicolumn{2}{c}{
\psfig{figure=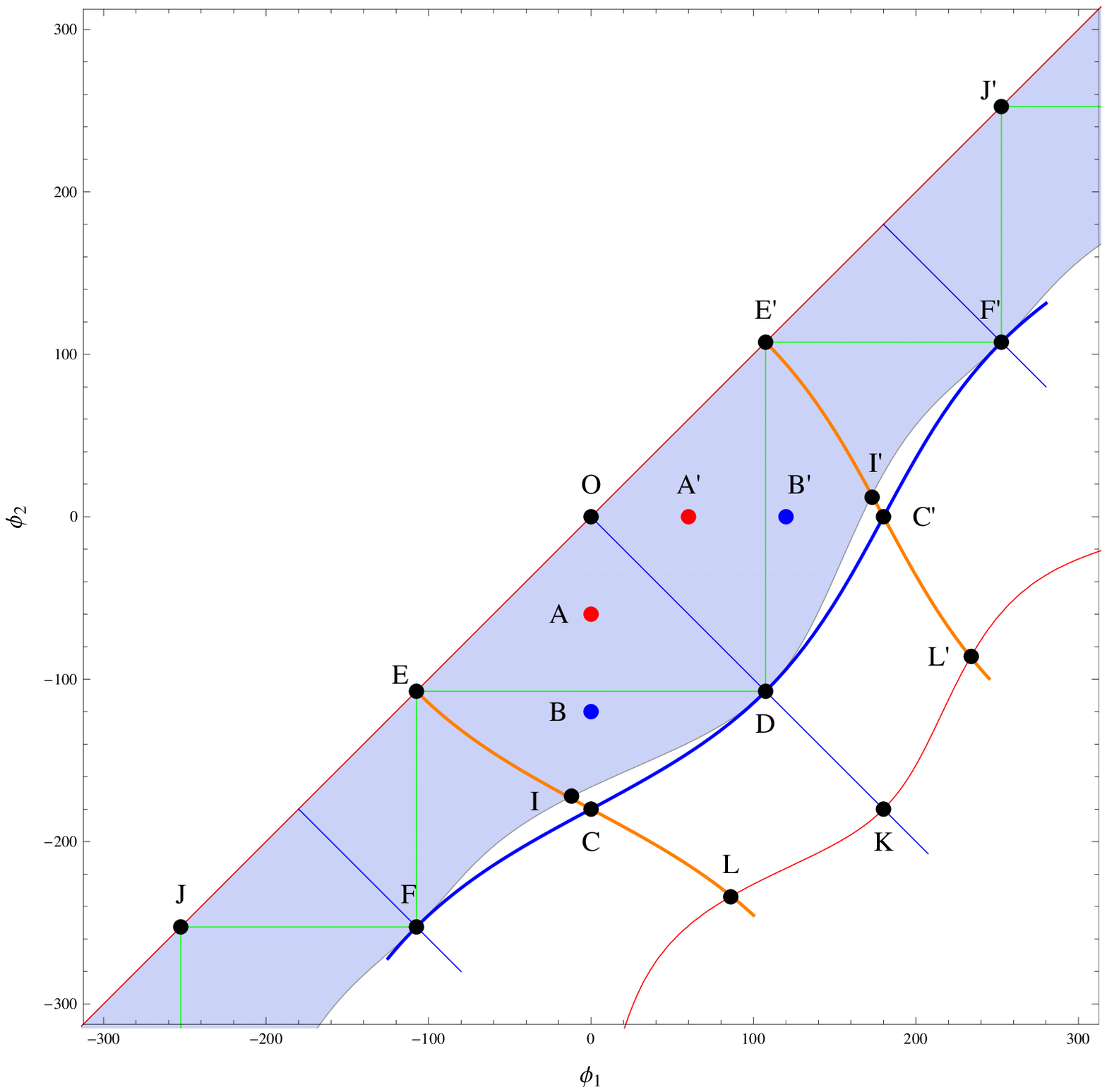,
height=5cm}} &
\psfig{figure=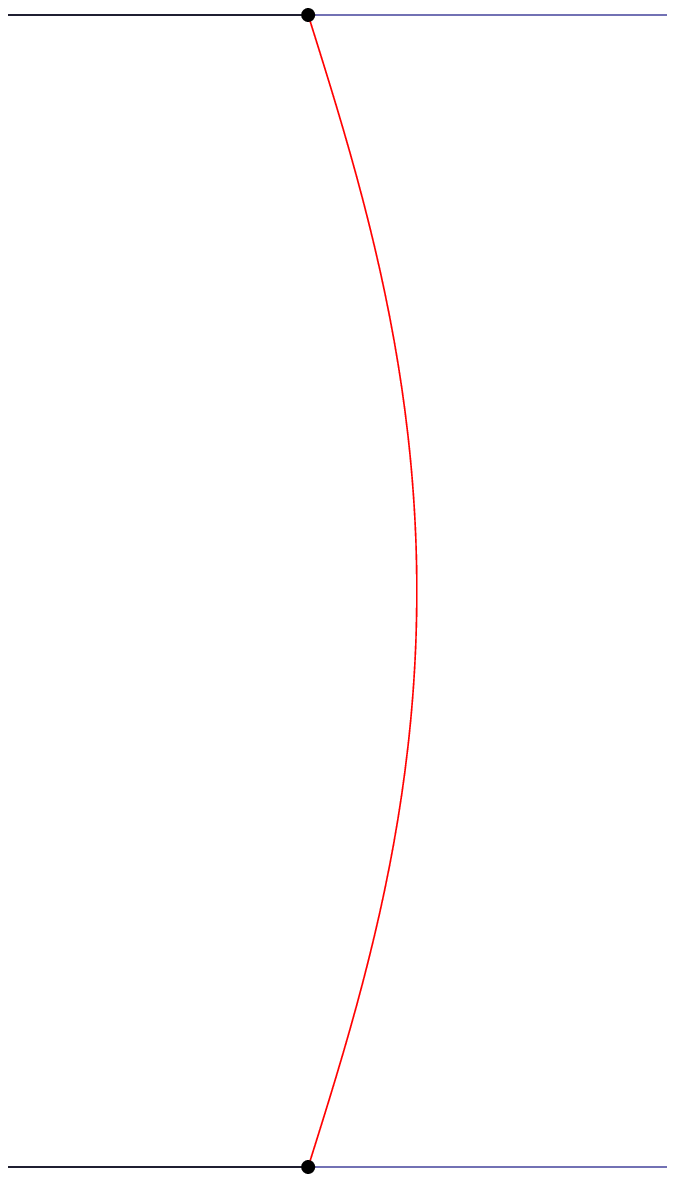,height=5cm}\\
\multicolumn{2}{c}{(a)} & (b) \\
\psfig{figure=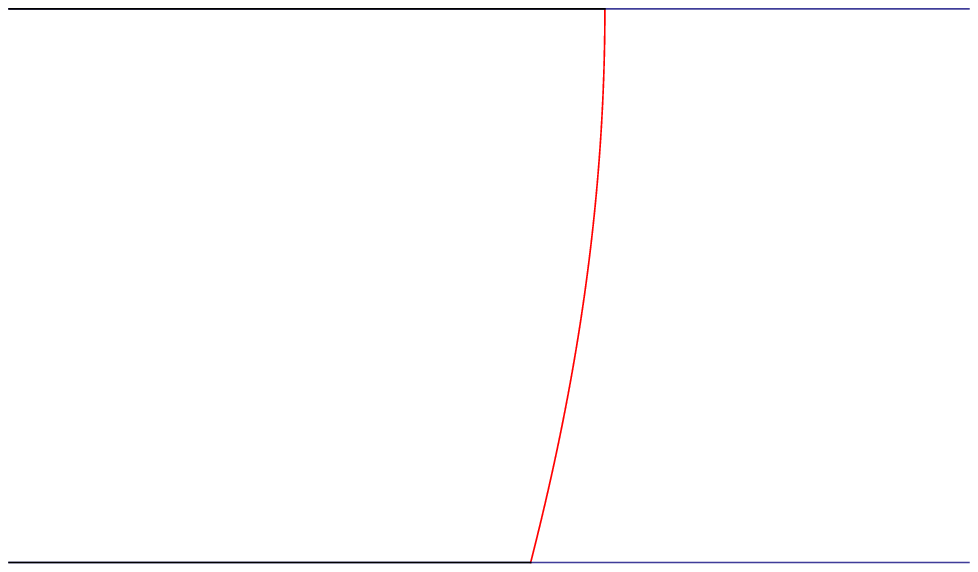,width=2.5cm}
\hspace{.7cm}&\hspace{.7cm}
\psfig{figure=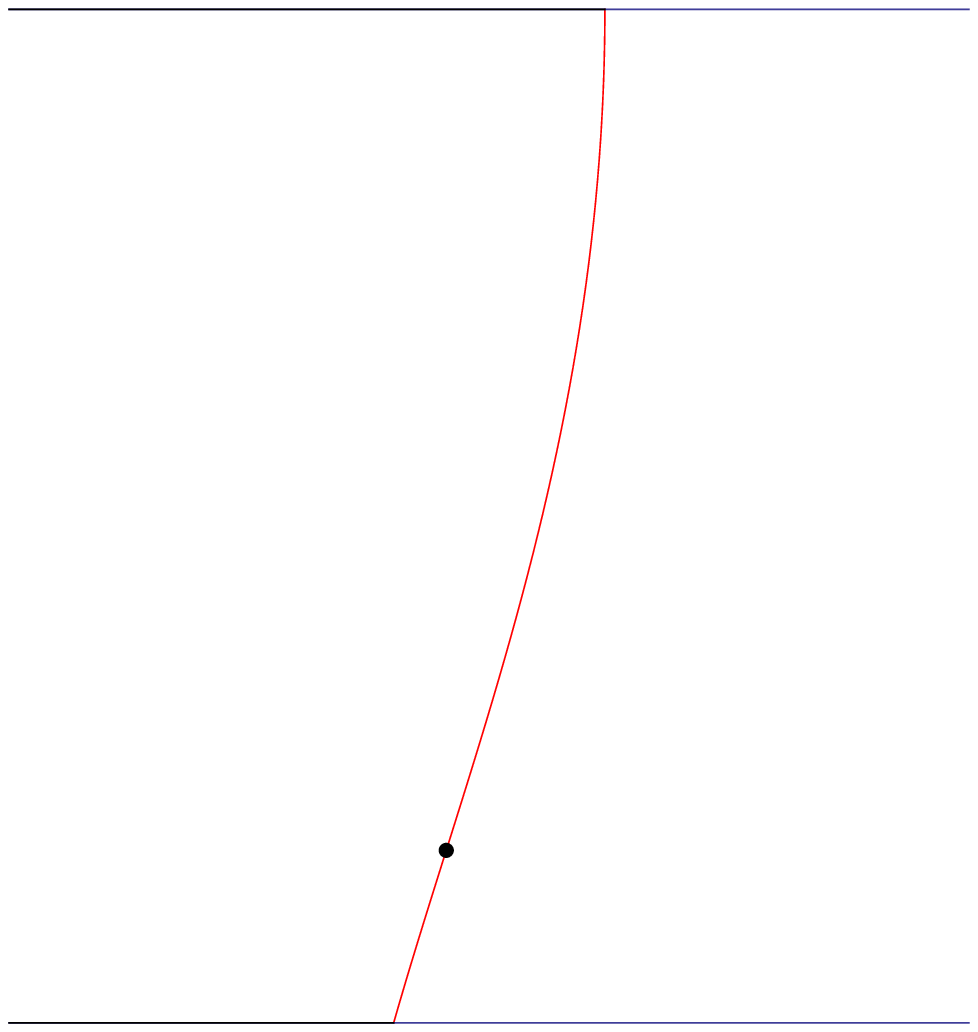,width=2.5cm}
\hspace{.7cm}&\hspace{.7cm}
\psfig{figure=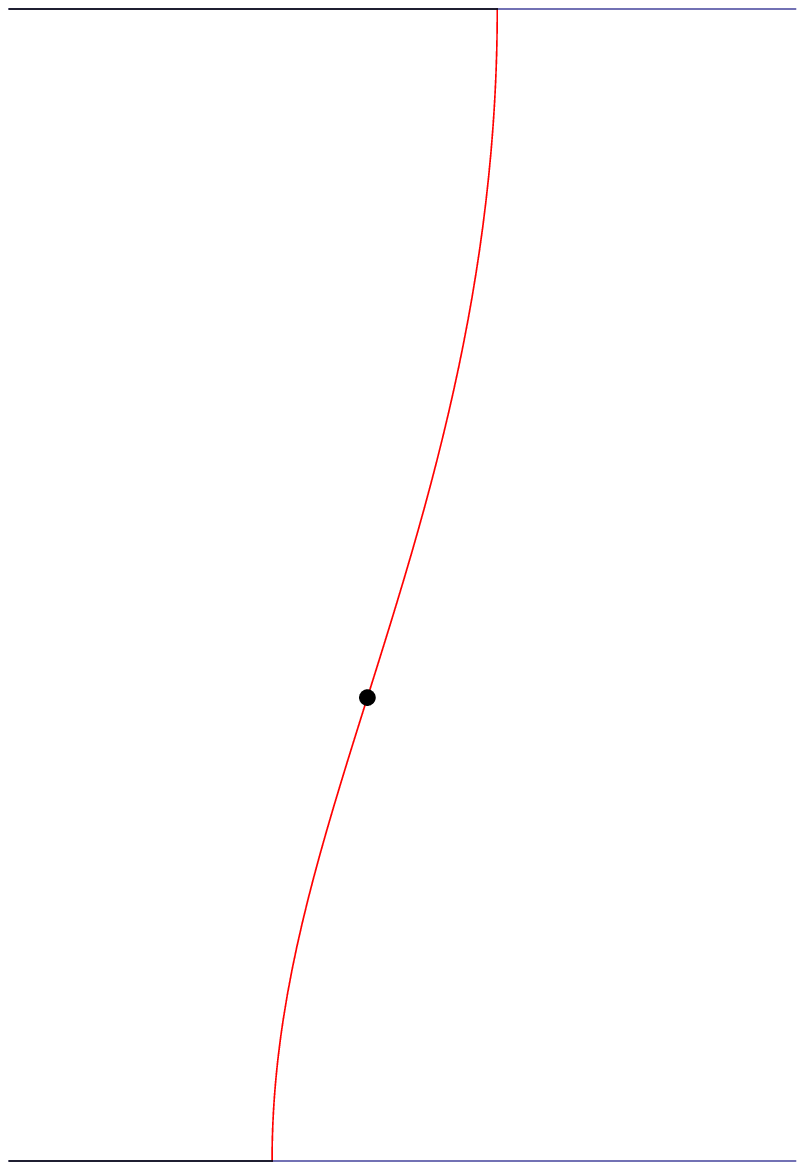,width=2.5cm}
\\
(c) & (d) & (e)
\end{tabular}\end{center}
\caption{The {\em SD} for ${\sf Und}$ menisci with $B=0.3$. The {\em green
lines} ${\mathfrak Z}$ show {\em IP} separation from a plate. The dots mark
menisci shown in: (b) point {\bf D} for $\phi_1=-\phi_2=\phi_U^{ip}=107.46^o$
(two {\em IPs} at the plates) (c) point {\bf A} for $\phi_1=0^o,\phi_2=-60^o$
(stable meniscus without {\em IP}), (d) point {\bf B} for $\phi_1=0^o,\phi_2=
-120^o$ (stable meniscus with one {\em IP}), (e) point {\bf C} for $\phi_1=0^o,
\phi_2=-180^o$ (unstable meniscus with one {\em IP}).}\label{green}
\end{figure}

\noindent
The ${\sf Und}$ menisci with such BC have necessarily one or more {\em IPs}: one
{\em IP} for $\phi_1=n\pi,\phi_2=(n-1)\pi$, two {\em IP}s if $\phi_1=n\pi,\phi_2
=(n-2)\pi$, {\em etc.}, where $n$ is an integer. However, for $n\geq 2$ a 
criterion (\ref{f12}) is broken, {\em i.e.}, the conjugate points appear. So
there remains one {\em IP} and a direct calculation of $Q_{33}$ gives for
$0<m<1$,
\bea
4\frac{Q_{33}(0,-\pi)}{(1-B)^2}\!=\![3E(m)-K(m)][E(m)-K(m)]+m^2K(m)[2E(m)-
K(m)]<0,\nonumber
\eea
where $K(m)$ and $E(m)$ denote the complete elliptic integral of the first and
second kind. The last inequality may be verified numerically. In Figure
\ref{green} we present detailed locations of ${\sf Und}$ menisci with $B=0.3$
in the sense of its stability w.r.t. the boundaries $\Delta(\phi_1,\phi_2)=0$ 
({\em the red curve} ${\mathfrak R}$) and $Q_{33}(\phi_1,\phi_2)=0$ ({\em the 
gray curve} ${\mathfrak G}$). The points $C(\phi_1=0,\phi_2=-\pi)$ and $C'
(\phi_1=\pi,\phi_2=0)$ lie in unstable zone.

{\bf 2}. {\em If $\;\theta_1=\theta_2$ there are no stable menisci with one or
more IPs \cite{FinVog1992}, Theorem 5.7}.

\noindent
All ${\sf Und}$ menisci with $\theta_1=\theta_2$ and without {\em IP} have the
endpoints satisfying $\phi_1+\phi_2=0$. In Figure \ref{green} they belong to the
interval OD of the {\em blue line} ${\mathfrak B}$ and are stable. There are
two different ways to generate {\it IP}.

First, allow $\phi_1$ to grow by preserving the above equality that leaves the
meniscus symmetric w.r.t. reflection plane between two plates. When $\phi_1=
\phi_U^{ip}$ there appears a couple of {\em IPs} (see Figure \ref{green}(b)),
{\em i.e.}, {\em IP}s are born on both plates simulateneously. We cannot make
any conclusion about stability of this meniscus in the framework of Weierstrass'
theory. But all menisci with $\phi_1+\phi_2=0$, $\phi_1>\phi_U^{ip}$, having two
{\em IPs} are unstable. In Figure \ref{green} they belong to ${\mathfrak S}$ 
beyond the point D. Thus, the range of equal contact angles $\theta$ for stable 
menisci without {\em IP} reads, $\pi/2<\theta<\phi_U^{ip}$ for convex ${\sf 
Und}$ and $\pi-\phi_U^{ip}<\theta<\pi/2$ for concave ${\sf Und}$.

Another way to generate {\em IP} with $\theta_1=\theta_2$ is to break the
reflection symmetry $\phi_1+\phi_2\neq 0,$ where $\phi_1\!<\!\phi_U^{ip}$ and
$\phi_2\!<\!-\phi_U^{ip}$. Using (\ref{k2}) for $\;\tan\theta_j\!=\!(-1)^{j-1}z'
(\phi_j)/r'(\phi_j)$ write an equality for $\phi_1,\phi_2$,
\bea
P(\phi_1)+P(\phi_2)=0,\quad P(\phi)=\frac{1+B\cos\phi}{B\sin\phi}\;\Rightarrow
\;\tan\frac{\phi_1}{2}\tan\frac{\phi_2}{2}=-\frac{1+B}{1-B}.\quad\label{k21}
\eea
Calculation of $Q_{33}$ in accordance to (\ref{g15}) and (\ref{k21}) leads to
a cumbersome expression. Instead of its analysis we present in Figure
\ref{green} the {\em blue curve} ${\mathfrak B}$ given by equation (\ref{k21}) 
and observe that ${\mathfrak B}$ always lies in instability zone, confirmed by 
numerical calculation of $Q_{33}$ for $0<B<1$. The curves ${\mathfrak B}$ and 
${\mathfrak G}$ are tangent at points F, D, H.

There is one more important conclusion: ${\sf Und}$ meniscus with reflection
symmetry ($\theta_1=\theta_2$) and fixed CL at two plates is stable even when
two {\em IPs} exist. This follows from an observation that an interval DK at
Figure \ref{green} is above the curve ${\mathfrak R}$. The point $K(\phi_1=\pi,
\phi_2=-\pi)$ marks unstable ${\sf Und}$ meniscus of entire period with four
{\em IPs} when two of them are separated from the plates.

{\bf 3}. {\em If $\;\theta_1,\theta_2\neq\pi/2$, $\;\theta_1+\theta_2=\pi$ there
are stable menisci of large volume that have IPs \cite{Vog1989}, Remark 3.2}.

\noindent
Making use of (\ref{k2}) and identity $\tan(\theta_1+\theta_2)=0$ write a 
relation for the angles $\phi_1,\phi_2$ valid for the arbitrary volume's value,
\be
P(\phi_1)-P(\phi_2)=0\quad\Rightarrow\quad\tan\frac{\phi_1}{2}\tan\frac{\phi_2}
{2}=\frac{1+B}{1-B}.\label{k22}
\ee
Similarly to the previous case consider in Figure \ref{green} the {\em brown
curves} given by equation (\ref{k22}) and observe that they always pass through
the point C and cross transversely the curve ${\mathfrak G}$ at point I which
separates the menisci in two families: stable with one {\em IP} (at interval
GI) and unstable (beyond the point I). Note that the stable menisci without
{\em IP} are forbidden. Regarding the claim {\em 'stable menisci of large volume
that have IPs'} we have found it incorrect. Indeed, the whole segment E'I
belongs to the stability region and it remains true when we approach the point
E', {\em i.e.}, when $\phi_2\to\phi_1$ that manifests volume decrease up to an
arbitrary small value. Therefore we make a statement slightly different: {\em if
$\theta_1+\theta_2=\pi$ then only menisci with a single IP are stable}.

Summarize the above results: the stability region ${\sf Stab}_2(\phi_1,\phi_2)$ 
of ${\sf Und}$ meniscus between two plates with free CL is represented in 
Figure \ref{green} by interior of domain decomposed in subdomains
\bea
{\sf Stab}_2(\phi_1,\phi_2)=\!\{DIFE\}_1\!\cup\!\{DI'F'E'\}_1\cup\;\{JEF\}_0
\!\cup\!\{EOE'DE\}_0\!\cup\!\{J'E'F'\}_0\quad\nonumber
\eea
where a subscript stands for a number of {\em IP} in stable meniscus.

\begin{figure}[h!]\begin{center}
\psfig{figure=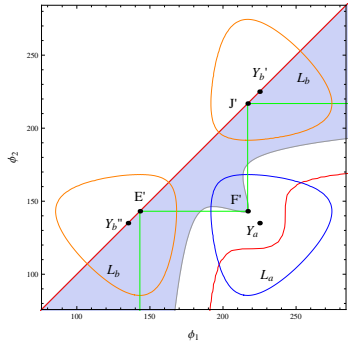,height=4.5cm}
\end{center}
\caption{The {\em SD} for ${\sf Und}$ menisci with $B=0.8$ (a part of Figure
\ref{C_S_U}(d)). The {\em green} line shows the position of {\em IPs}. The dots
mark menisci contacting the lower and upper plates at $\phi_2$ and $\phi_1$,
respectively: ($Y_a$) $\phi_1=225^o,\phi_2=135^o$, ($Y_b'$) $\phi_1=\phi_2=
225^o$, ($Y_b''$) $\phi_1=\phi_2=135^o$, in accordance with section \ref{s621}.
Curves ${L}_a$ and ${L}_b$ describe ${\sf Und}$ menisci satisfying $\theta_1+
\theta_2=\pi/2$ and $\theta_1-\theta_2=\pi/2$, respectively.}\label{blue}
\end{figure}

{\bf 4}. Finish this section with two other setups for ${\sf Und}$ menisci
between two plates: $\theta_1\pm\theta_2=\pi/2$, which differ from those
discussed in \cite{Ath1987}, \cite{Vog1987}, \cite{Vog1989}, \cite{FinVog1992}.
Making use of formulas (\ref{k2}) write an equality which is not solvable in
$\phi_1,\phi_2$ for all $B$,
\bea
P(\phi_1)P(\phi_2)=\mp 1,\;\left|P(\phi)\right|\geq\left|P\left(\phi_U^{ip}
\right)\right|=\frac{\sqrt{1-B^2}}{B},\;\Rightarrow\;\exists\;\;\phi_j,\in\Re
\;\;\mbox{if}\;\;B\geq\frac1{\sqrt{2}}.\nonumber
\eea
The upper (lower) sign in above equality corresponds to the upper (lower) sign
in $\theta_1\pm\theta_2$. For $B=1/\sqrt{2}$ there exist two pointwise solutions
of equation $P(\phi_1)P(\phi_2)=\mp 1$,
\bea
a):\;\frac{\phi_1}{5}\!=\!\frac{\phi_2}{3}\!=\!\frac{\pi}{4}\;\;\mbox{and}\;\;
\frac{\phi_1}{5}\!=\!\frac{\phi_2}{3}\!=\!-\frac{\pi}{4};\quad
b):\;\frac{\phi_1}{5}\!=\!\frac{\phi_2}{5}\!=\!\pm\frac{\pi}{4}\;\;\mbox{and}\;
\;\frac{\phi_1}{3}\!=\!\frac{\phi_2}{3}\!=\!\pm\frac{\pi}{4}.\nonumber
\eea
However, when $B>1/\sqrt{2}$ the solutions are represented by curves ${L}_a$
and ${L}_b$ in the halfplane $\{\phi_2<\phi_1\}$: ${L}_a$ passes through
unstable and stable (without {\em IP}) zones while ${L}_b$ exists only in stable
zones with and without {\em IP} (see Figure \ref{blue}).
\section*{Acknowledgement}
The useful discussions with O. Lavrenteva are appreciated. The research was
supported in part (LGF) by the Kamea Fellowship.

\end{document}